\documentclass{article}
\usepackage{graphicx} 
\usepackage{blindtext}
\usepackage[authoryear,round]{natbib}
\usepackage{amsmath,amssymb,amsfonts}
\usepackage{fancyhdr}
\usepackage{titlesec}
\usepackage{enumitem}
\usepackage{graphicx}
\usepackage{tikz}
\usepackage{cancel}
\usepackage{graphicx} 
\usepackage{appendix}
\usepackage{epsf}
\usepackage{amsmath}
\usepackage{here}
\usepackage{setspace}

\usepackage{varioref}
 \usepackage{wrapfig}
 \usepackage{threeparttable}
 \usepackage{dcolumn}
  \newcolumntype{d}{D{.}{.}{-1}}
 \usepackage{nomencl}
  \makeglossary
  \usepackage{subfigure}

\usepackage{subfigure}
 \usepackage{subfigmat}
 \usepackage{fancyvrb}
  \fvset{fontsize=\footnotesize,xleftmargin=2em}
 \usepackage{lettrine}
  
 \usepackage{subfigmat}
 \usepackage{fancyvrb}
 \usepackage{lettrine}

\usepackage{caption}
\usepackage{epstopdf}
\usepackage{lineno}
\usepackage[margin=1in]{geometry}
\usepackage{graphicx} 
\usepackage{epstopdf}
\usepackage[caption=false]{subfig}

\renewcommand\bibfont{\fontsize{10}{12}\selectfont}
\usepackage{epstopdf}
\usepackage[caption=false]{subfig}

\usepackage{bigints}

\renewcommand\bibfont{\fontsize{10}{12}\selectfont}

\usepackage{multirow}

\begin{document}


\title{Perturbation Theory for Translating Oblate-Spheroidal Droplets with Internal Circulation}
\author{William A. Sirignano 
\\ University of California, Irvine}

\date{September 24, 2026}

\maketitle

\begin{abstract}
Liquid droplets deform from spherical shape due to aerodynamic variation of pressure along the surface as the droplet moves through a gas. The deformation is predicted for axisymmetric droplets translating through a gas with low Weber numbers, $We < 1$, and Reynolds number $Re = O(10)$. That deformation analysis is based on the relations between local pressure jump and the two radii of curvature. A thin boundary layer on both sides of the gas-liquid interface is considered with a surface-velocity jump due to pressure-gradient-driven flow with a large density jump and a pressure jump due to surface tension. A near-ellipsoidal shape is predicted using $We$ as a perturbation parameter. Then, the quasi-steady internal liquid-phase stream function and velocity field are predicted, describing internal circulation and a vortex ring structure with vorticity distributed through an inviscid liquid. The gas-phase flow over the oblate droplet is described using a ring doublet as an image within the droplet. The ring-doublet radius is related to $We$. Gas potential flow results are presented and compared using both the exact analytical solution and a perturbation analysis based on $\sqrt{We}$. The perturbation analysis provides a lower computational cost.     Three analyses for local curvature, liquid circulation, and gas potential flow are matched to yield the  velocity and pressure fields.
The appropriate radius for the image ring doublet is matched to $\sqrt{We}$. Stream functions and the two velocity components in both fluids and gas potential field are predicted. Some comments on droplet drag are presented.
\end{abstract}

\index{keyword} Keywords: oblate droplet flow, internal circulation, ring-doublet image

\section{Introduction}

For decades, there has been interest in the deformation and break-up of falling or translating droplets and rising bubbles. The problem brings together aerodynamic behavior and surface-tension effects. Interest in droplets and bubbles has led to examination of configurations with higher relative velocities and shear forces, thereby  with increased deformation and greater likelihood of breakup. As translational velocity increases and/or surface tension decreases, the deformation increases and the likelihood for breakup of the droplet increases. We will focus on a particular range of Reynolds number and Weber number and not attempt to review a very broad literature. The interested reader can examine certain publications to understand better the wide interest: \cite{Stone}, \cite{Magnaudet}, \cite{Dai}, \cite{Jackiw}, \cite{Szakall}, and \cite{Dabiri}.

There are similarities in the gravity-driven behaviors of the rising bubble and the falling droplet although parameter magnitudes differ greatly.  Early interest on droplet and bubble deformation focused on slow gravity-driven motion of falling droplets or rising bubbles. In those cases, both the Weber number $We$ and the Reynolds number $Re$ could be small and the boundary-layer thickness in the carrier fluid was of comparable or larger size than the droplet or bubble measure. This resulted in treatments using Stokes flow to study the field around the oblate spheroid. \cite{TaylorAcrivos} used perturbation theory with $Re < 1$ to study the flow around and inside the falling droplet. \cite{Moore} explains how distortion of the form of a  rising bubble from a sphere to an oblate spheroid depends on $We$ for small values of that parameter; The ratio of the semi-major axis to the semi-minor axis is approximated by $1 + 9 We /64 +O(We^2)$.

In many applications, especially some related to power and propulsion, we find a range of parameters where Weber number $We \equiv 2 \rho_g U_{\infty}^2 R_0/\gamma \leq 1$ and Reynolds number $Re \equiv 2 \rho_g U_{\infty}R_0/\mu_g > 10$. Here, $\rho_g, U_{\infty}, R_0, \gamma$ and $\mu_g$ are the gas density, free-stream gas velocity in a reference frame moving with the droplet,  droplet radius if the given volume were in a spherical shape, coefficient of surface tension, and coefficient of dynamic viscosity, respectively. For example, a decane droplet with diameter about $50\;\mu m$ moving at velocity $U_{\infty} = 5\; m/s$ through  air at pressure $p= 4 \;bar$ approximately results in $We = 1$ and $Re = 50$. The study here focuses on this range where $We<1$ but $10 < Re < 25$ where a thin gas-phase boundary layer exists with surrounding irrotational gas flow and internal liquid circulation but without a recirculating gas wake. Quasi-steady flow is addressed through a steady-state analysis. We do not address issues related to droplet oscillation, breakup, or non-axisymmetric behavior, all of which occur at higher $We$ values.  Surface tension remains constant over the surface; effects of temperature variation and/or  additives on surface tension are not considered. An aim is to build a foundation for later studies on droplet heating and vaporization and/ or on droplet drag and deceleration. Consideration is given to formulations that would be computationally inexpensive for use in sub-grid modelling of spray flows.

We present and coordinate three analyses. Variation in pressure and its impact on droplet curvature variation and droplet shape is discussed in Section \ref{curve}. Internal circulation is analyzed in Section \ref{internal}  while the gas potential flow is addressed in Section   \ref{potent}. 
The integration of these components appears in Section \ref{integrate} followed by concluding remarks in Section \ref{conclude}.

\section{Surface Curvature} \label{curve}

Consider a cylindrical coordinate system for the axisymmetric description with the origin at the droplet center; $r$ is the radial direction, and $x$ is the axial direction. The droplet surface is described by the function $g(x)$. Define $g' = dg/dx$ and $g'' = d^2 g/dx^2$. The curvature is described by the meridional curvature radius $R_{\phi} \equiv [1 + (g')^2]^{3/2}/|g''|$ and the circumferential curvature radius $R_{\theta} \equiv g\sqrt{1 + (g')^2}$. For a sphere, each of these  curvature radii equals the sphere radius. For an oblate spheroid, they will differ from each other. We assume thin boundary layers on both sides of the phase interface. Thus, the only cause of a pressure jump is the surface tension. The pressure jump across the phase interface is given by the following relation.

\begin{eqnarray}
 \Delta p = \gamma \Big[\frac{1}{R_{\theta}} + \frac{1}{R_{\phi}}\Big]
= \gamma \Big[\frac{1}{g\sqrt{1 + (g')^2}}  + \frac{|g''|} {[1 + (g')^2]^{3/2}}  \Big]
    \label{curvature}
\end{eqnarray}
Caution is needed with the singularity on the axis of the spheroid where $g=0$ and $g' \rightarrow \infty$.

Still, the pressure on each side of the interface will vary along the interface due to the varying tangential velocity and the Bernoulli effect. Thereby,  at the interface
\begin{eqnarray}
    p_g = p_{g, stag} - \rho_g U_{g,s}^2/2  \nonumber \\
    p_l = p_{l, stag} - \rho_l U_{l,s}^2/2
\end{eqnarray}
The subscripts $g, l , s,$ and  $stag$ imply gas, liquid, surface, and stagnation point, respectively. The corrected Weber number will be defined  on the basis of the difference in the maximum inertia of the two fluids at the surface for a spherical droplet
\begin{eqnarray}
     We_{corrected} = \frac{2 R_0[\rho_{g,s}  U_{g,S,max}^2 - \rho_{l,s} U_{l,S,max}^2]}{\gamma}
\end{eqnarray}
Specifically, we first take the pressure difference given along a spherical droplet. Next, we will determine the shape change caused by that small variation of pressure along the surface. The literature shows that tangential surface velocity along the sphere has $U_{g,s, max} = (3/2) U_{\infty},  U_{g,s} = U_{g,s, max} sin \theta = (3/2) U_{\infty} sin \theta   , 
U_{l,s} = U_{l,s, max} sin \theta,$ where $sin \theta = x/\sqrt{x^2 +r^2}$.
Then,
\begin{eqnarray}
\frac{\Delta p}{\gamma} - \frac{\Delta p_{stag}}{\gamma} =\frac{We_{corrected}}{4} sin^2 \theta = \frac{We_{corrected}}{4}\Big(\frac{x^2}{r^2 + x^2}\Big)
\end{eqnarray}
The difference in the stagnation pressures will be taken as the affect of surface tension with the original spherical radius, $\Delta p_{stag} = 2 R_0/ \gamma$.
We will have this pressure difference along the surface affect the local curvature and the droplet shape. For the small Weber number, we define the small perturbation parameter as $\delta \equiv We_{corrected}/ 4.$ From this point forward, $We$ will refer to the corrected Weber number as defined earlier.   The result follows:
\begin{eqnarray}
  \frac{1}{g\sqrt{1 + (g')^2}}  + \frac{|g''|} {[1 + (g')^2]^{3/2}}                      = \frac{2R_0}{\gamma} + \delta\Big(\frac{x^2}{r^2 + x^2}\Big)
    \label{curvature2}
\end{eqnarray}

    We can solve Equation (\ref{curvature2}) using an expansion in the small positive parameter $\delta$; namely, $r = g(x) = g_0(x) + \delta g_1(x) + \delta^2g_2(x) + O(\delta^3)$ 
    where the leading term is the spherical description $g_0(x) = \sqrt{R_0^2 - x^2}$.
    We normalize the variables for convenience, yielding $X= x/R_0, G_0 = g_0/R_0,$ and $G_n = g_n/R_0$ for $n = 1, 2, ...$. Solution is pursued only through the first-order term $g_1(x)$; we will allow quantitative error of $O(\delta^2)$.
    Substitution of the series and separation of coefficients of $\delta^n$ under the principle of linear independence yields the solution for $G_0$ and a linear second-order non-homogeneous ordinary differential equation for $G_1$.
\begin{eqnarray}
    &&G_0 = (1 - X^2)^{1/2}  \\
    &&G_0^4\frac{d^2 G_1}{dX^2} -  4 XG_0^2 \frac{dG_1}{dX} + G_1 = - G_0^3 = - (1 - X^2)^{3/2}
    \label{gode}
\end{eqnarray}
Since $X^2$ does not exceed the magnitude of unity in the liquid droplet, the nonhomogeneous term may be expanded in a power series of $X^2$ terms. Accordingly, $G_1$ may be expanded in a similar $X^2$ powers. This feature reflects that the solution for surface value of cylindrical radius is symmetric in $x$. Actually, our analysis can simplify by using the squares of both the independent variable and the dependent variable in the above ordinary differential equation \ref{gode}. Specifically, we set
\begin{eqnarray}
\eta \equiv G^2 = G_0^2 + 2\delta G_0G_1 + \delta^2[G_1^2 + 2G_0G_2] + O(\delta^3) \;\; \;\;; \;\; \;\;  \xi \equiv X^2
\end{eqnarray}
$H_0 \equiv G_0^2 = 1 -\xi$ readily follows and, defining $H_1= 2G_0G_1$ , we obtain the new form of the ordinary differential equation
\begin{eqnarray}
\xi(\xi -1)\frac{d^2 H_1}{d\xi^2} +  \frac{(1 - 3\xi)}{2} \frac{dH_1}{d\xi} + \frac{1}{2}H_1 = \frac{\xi -1}{2}  
    \label{hode}
\end{eqnarray}

The particular solution to Equation (\ref{hode}) is linear in $\xi$ and can readily be found to be $H_{1, particular} = -(\xi+1)/2$. We will use series expansions to obtain the homogeneous solution. Both $\xi = 0 $ and $\xi =1$ are regular singular points with some guarantee of a series solution. We expect the semi-minor axis to be smaller than the original spherical radius; thus, $0 \leq \xi < 1$ along the droplet surface. We use expansion about $\xi = 0$. Two solution forms are found with $H_{1,homogeneous} = \Sigma_{m=0}^{\infty}a_m\xi^m  + \xi^{1/2}\Sigma_{m=0}^{\infty}b_m\xi^m $.  Since $G_1 = H_1/(2G_0)= (H_{1,homogeneous} +
H_{1, particular} )/(2G_0)$ should have zero slope at $x=0$, i.e., $\xi =0$, we dismiss the series that is not analytic at $\xi =0)$, setting all $b_m =0$ and keeping $H_1 =  -(\xi + 1)/2  +  \Sigma_{m=0}^{\infty}a_m\xi^m $. Now, the coefficients $a_m$ remain to be determined.

Substitution of the analytic series into the homogeneous form of Equation (\ref{hode}) yields
\begin{eqnarray}
    a_{m+1} = \frac{2m^2 +m -1}{(2m+1)(m+1)}a_m = \frac{2m -1}{2m+1}a_m  \;\;\; ; \;\;\;m = 0, 1, 2, .....
    \label{coeff}
\end{eqnarray}
For large $m$ values, $a_{m+1}/a_m \rightarrow 1 $.  The coefficient values for $m\geq 1$ decrease with increasing $m$ and all have opposite sign to $a_0$. The general result is $a_m =- a_0/(2m-1)$. For example, $ a_1/a_0 =-1 ;  a_2 =(1/3)a_1 = -a_0/3;  a_3 =(3/5)a_2 = - a_0/5  ;  a_4 =(5/7)a_3 = -a_0/7, a_5= (7/9) a_4 = - a_0/9$. The solution at this point is given by
\begin{eqnarray}
    H(\xi) = H_0 + \delta H_1 +O(\delta^2) = 1-\xi + \delta \Big[- \frac{\xi +1}{2} + a_0 \Big( 1  -   \Sigma_{m=1}^{\infty} \frac{\xi^m}{2m-1} \Big)  \Big]  + O(\delta^2)
    \label{H1}
\end{eqnarray}

A simplification can follow by relating the infinite series to a specific function using X as the independent variable. Note that, for $0 \leq X < 1$, we may use a Taylor series and set 
\begin{eqnarray}
    \frac{1}{1 -X^2} = \Sigma_{m=0}^{\infty} X^{2m} 
\end{eqnarray}
Thus, the following pathway is followed.
\begin{eqnarray}
F(X) &\equiv& \Big[\Sigma_{m=1}^{\infty}\ \frac{X^{2m}}{2m-1} \Big] = X \Big[\Sigma_{m=1}^{\infty} \frac{X^{2m-1}}{2m-1} \Big] \;\; ; \;\;\nonumber \\
\frac{d}{dX}\Big[\frac{F}{X}\Big] &=&  \frac{d}{dX}\Big[\Sigma_{m=1}^{\infty}\ \frac{X^{2m-1}}{2m-1} \Big]  =  \Sigma_{m=1}^{\infty} X^{2m-2} = \frac{1}{X^2}\Big[\frac{1}{1-X^2}-1\Big]  \;\; ; \;\;\nonumber \\
\frac{F}{X} &=& A + \int_0^X \frac{d X'}{(X')^2(1 - (X')^2)} - \int_0^X \frac{dX'}{(X')^2} = A + \int_0^X \frac{d X'}{(1 - (X')^2)}
\end{eqnarray}
By the above definition of $F(X)$, the ratio $F/X$ must be zero at $X =0$ and thereby  the constant $A=0$, yielding
\begin{eqnarray}
F &=& -\frac{X}{2} ln(1-X^2) = -\sqrt{\xi}\;ln(\sqrt{1-\xi}) = -ln (G_0)
\end{eqnarray}
Only positive $X$ values need be considered, given the symmetry.  Note that $F(X) - X = -\frac{X}{2} ln(1-X^2) -X = O(X^4)$.   Now, 
\begin{eqnarray}
    H(\xi)  &=& 1-\xi + \delta \Big[- \frac{\xi +1}{2} + a_0 \Big( 1  +  \sqrt{\xi}\; ln(\sqrt{1-\xi}\Big)  \Big]  + O(\delta^2)  \nonumber \\
    &=& 1-\xi + \delta \Big[- \frac{\xi +1}{2} + a_0 \Big( 1  -\xi +  \big[\sqrt{\xi}\; ln(\sqrt{1-\xi} +\xi \big]\Big) \Big]  + O(\delta^2) \nonumber \\
    &=& 1-\xi + \delta \Big[- \frac{\xi +1}{2} + a_0 \Big( 1  -\xi \Big) \Big] +O(\delta \xi^2) + O(\delta^2) 
    \label{H2}
\end{eqnarray}

The logarithmic term causes a deviation of $O(\delta |X|^4)$ from the ellipsoidal shape. The term is zero at  $\xi=0$, but grows and reduces the semi-minor axis compared to an ellipse. The droplet shape is described by $H = G^2 =  1 + \delta(a_0 -1/2) - (1 +\delta/2 +a_0\delta)X^2 - a_0 \;\delta (X/2 \;ln(1-X^2) +X^2 )  +O(\delta^2)$.

A remaining task is to determine the constant $a_0$. The deformation from the spherical shape will maintain the original volume of the sphere at the non-dimensional value of $\int_{-1}^{1} \pi H dX = \int_{-1}^{1} \pi G^2 = 4 \pi /3$. Accounting for symmetry, we only need to integrate from $X = 0$ to $X=1$ and double the value.  With zero change in volume as the shape changes, we have $\int_0^1 H_1 dX = 0$ as the condition to determine $a_0$. Using that condition together with integration of Equation (\ref{H1}) yields
\begin{eqnarray}
    a_0 = \frac {2/3}{1 - \Sigma_{m=1}^{\infty} [1/(4m^2 -1)]} = \frac {2/3}{1 -1/2} = \frac{4}{3}
    \label{a0}
\end{eqnarray}
 Now, we will consider the following result as our most accurate representation after neglect of terms of $O(\delta^2)$.
\begin{eqnarray}
 H(X)  = G(X)^2 = 1-X^2 + \delta \Big[- \frac{X^2}{2} + \frac{5}{6} + \frac{2}{3}|X|\;ln(1-X^2)\Big)  \Big]  + O(\delta^2) 
  \label{G}
\end{eqnarray}
The  above formulation carries an error of $O(\delta^2)$ because the integral of $G_1$ should terminate at a value Of $\xi <1$ because the semi-minor axis is shortened.

If $\xi^*$ is the value where radius $G=0$ and its square $H= 0$ through first order in $\delta$, we have from Equations (\ref{H2}) and (\ref{a0})
\begin{eqnarray}
   \Big(1 + \frac{11 \delta}{6}  \Big) \xi^* = 1 + \frac{5\delta}{6}   + \frac{4\delta}{3} \Big[\frac{\sqrt{\xi^*}}{2} \;ln (1 - \xi^*) +\xi^*\Big] +O(\delta^2)
   \label{xi}
\end{eqnarray}

We seek a  solution in the form $\xi^* = 1 - \beta\delta + O(\delta^2)$;  The corresponding value is $X^* = 1 - \beta \delta/2 +O(\delta^2)$.
From substitution into Equation (\ref{xi}), $\beta = -1/3 - (2/3) (1- \delta \beta/2)  \;ln(\beta\delta ) +O(\delta^2) $.   $\beta$ does depend on $\delta$.  Note that $\delta \;ln \;\delta \rightarrow 0$ as $\delta \rightarrow 0$.
We may solve this relation for $\beta$ by simple arithmetic iteration. As examples calculated roughly to two significant digits and within $O(\delta^2)$, with $We = 0.4$ and thereby $\delta = 0.1$, we have  $\xi^*=0.89$, and $X^* = 0.94$; with $We =0.8$ and $\delta = 0.2$, we find $\xi^* = 0.84$, and $X^* = 0.92$. (Our plots that follow were done with greater accuracy.)

From Equation (\ref{H2}) with substitution for 
$a_0$ and return to $G$ and $X$ variables, the near-ellipsoidal form can be developed by simple algebra and proper ordering in terms of the perturbation series.
\begin{eqnarray}
1 +\frac{5\delta}{6} = G^2 + \Big(1 +\frac{11\delta}{6}\Big)X^2  + O(\delta |X|^4) + O(\delta^2)   \nonumber \\
1 = \frac{G^2}{(1 + 5\delta/ 12)^2} + \frac{X^2}{(1 -\delta/2)^2}  + O(\delta |X|^4) + O(\delta^2)
\label{Gapprox}
\end{eqnarray}
The semi-major axis $a = 1 +5 \delta/12$ is larger than the original sphere radius while the semi-minor axis $ b = 1- \delta/2$ is smaller, as expected. It is seen that neglect of the $O(X^4)$ impact would predict $\beta = 1/2$ which is not consistent with the analysis for the $\beta$ value. The following approximation might be more useful.
\begin{eqnarray}
    1 \approx \frac{G^2}{(1 + 5\delta/ 12)^2} + \frac{X^2}{(1 -5\delta /6)^2}
    \label{Gbeta}
\end{eqnarray} 
Here, the two-to-one ratio of the perturbations for the semi-minor axis and semi-major axis is consistent with retaining the original volume of a sphere: namely, $(4\pi/3)(1+5\delta/12)^2(1 -5\delta) = (4\pi/3)(1^3 +O(\delta^2))$.

\begin{figure}[thbp]
\centering
 \subfigure [Droplet shapes for $\delta = 0.1$.]{
  \includegraphics[height = 5.6cm]{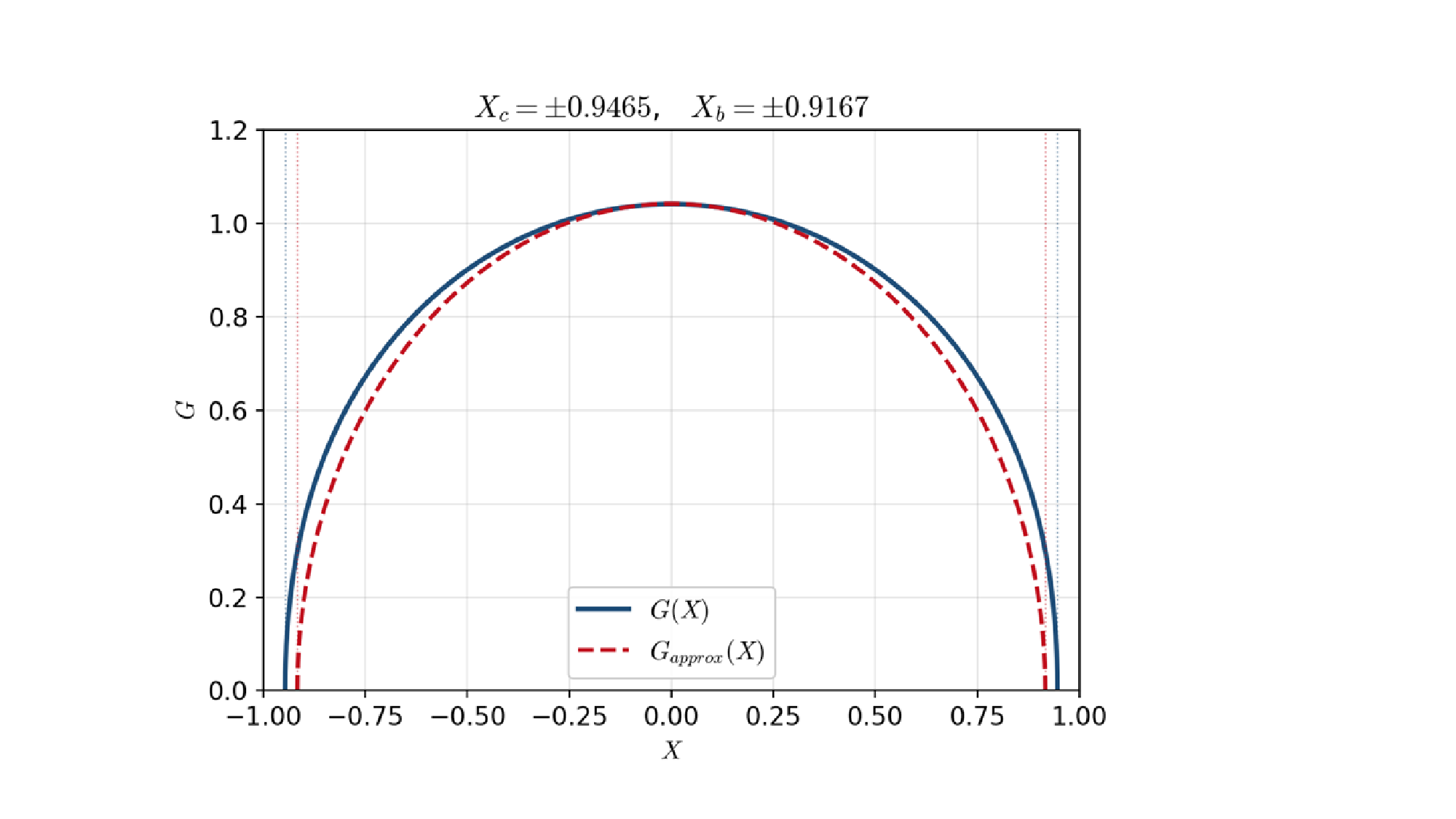}}  \\
  \centering
  \subfigure [Droplet shapes for $\delta = 0.2$ . ]{
  \includegraphics[height = 5.6cm]{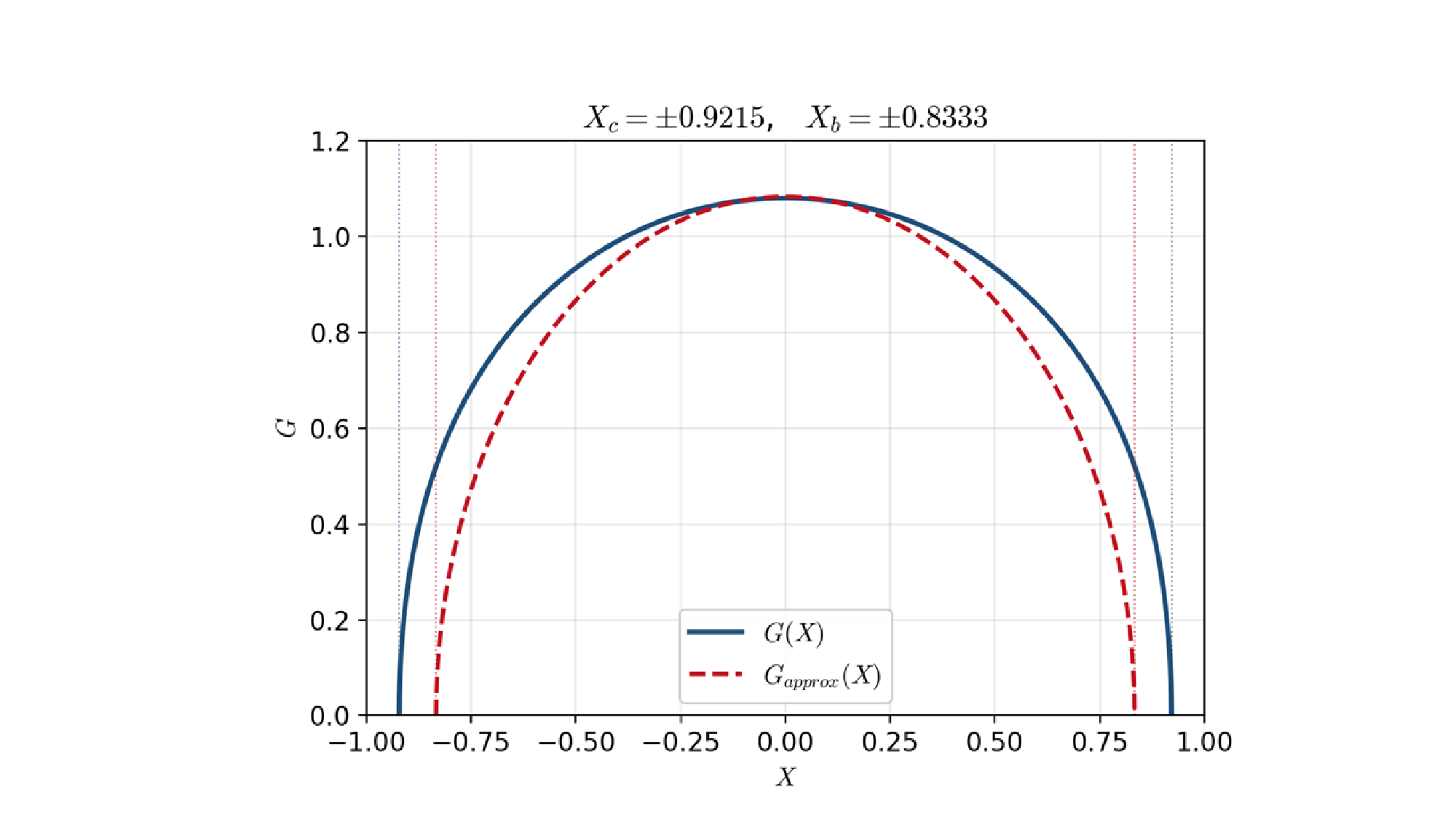}}
   \caption{Comparison of droplet shapes for results of $G_{approx}$ from Equation (\ref{Gbeta}) and $G$ from  Equation (\ref{G})
    for $\delta = 0.1 \;(We = 0.4)$ and $\delta = 0.2\; (We = 0.8)$. }
  \label{Surface1}
  \end{figure}
In Figure \ref{Surface1}, we compare the results from the ellipsoidal approximation given by Equation (\ref{Gbeta}) with the more exact formation of Equation (\ref{G}) which retains terms to all powers of $X^2$.  It is seen that the elliptical profiles gives a decrease in both semi-minor axis, while retaining the same semi-major axis. The value of $\beta =5/3$ preserves droplet volume within $O(\delta^2)$ accuracy without using the $O(X^4)$ and higher-order terms. It is noted that, with Equation (\ref{G}), the magnitude of the perturbation to the semi-major axis is very close to twice the perturbation magnitude of the semi-minor axis, but the two perturbations have opposite signs. This gives some justification for the choice for $\beta =5/3$ here which approximately matches the semi-minor axes for the fixed semi-major axis. However, realize that the analysis has an error of $O(\delta^2)$, which for $\delta = 0.2$ places the exercise in an error range that approaches 10 percent. However, the error is only significant near the stagnation points at low values of the cylindrical radius. The variation of liquid per segment of  axial distance is small there;  the axial-oriented  liquid volume for  an ellipsoid (or sphere) where the surface position has $r< a/2, |x| > \sqrt{3}b/2$ is $(4\pi a^2b/3)(1-9\sqrt{3}/16)$ which is $2.57$ per cent of the total liquid volume.

Using the ellipsoidal shape dictated by Equation (\ref{Gbeta}) with $a \equiv 1 +5 \delta /12$ and $b \equiv 1-5\delta /6$, Equation (\ref{curvature}) yields for pressure increase by jump across the surface
\begin{eqnarray}
    \Delta p = \frac{\gamma}{a}\Bigg[\frac{1 + \frac{a^2}{b^2}  -\frac{X^2}{b^2}          +\frac{a^2X^2}{b^4} } {\big( 1 - \frac{X^2}{b^2}  + \frac{a^2X^2}{b^4}   \big)^{3/2}}  \Bigg]
    \label{pjump}
\end{eqnarray}
Using Bernoulli's equation along each side of the droplet surface to relate pressure variation and velocity variation together with Equation (\ref{pjump}), the relation between maximum liquid velocity and maximum gas velocity is obtained. Of course, the maxima occur at $X=0, G=a$.
\begin{eqnarray}
U_{l, max} = \sqrt{\frac{\rho_g}{\rho_{l}}U_{g,max}^2  +\frac{2 \gamma}{a\rho_l}\big( 1- \frac{b}{a}\big)^2}  = \sqrt{\frac{\rho_g}{\rho_l}}U_{g,max} + O(\delta^2)
\label{maxvelocity}
\end{eqnarray}
So, the ellipsoidal shape does not allow an $O(\delta)$ impact on the maximum surface velocity.  Given that the liquid density will be two-to-three orders of magnitude larger than gas density in many applications, the liquid velocity will be very small compared to gas velocity. Friction will result from the velocity jump across the surface and it will contribute to the drag. However, friction is not driving liquid motion; rather, pressure gradient drives it while friction is created from the viscous opposition to being driven.

\section{Internal Liquid Circulation} \label{internal}

The internal circulation of the deformed droplet will be examined. It is important in the calculation of the pressure jump across the interface with the gas; thereby, it affects local surface curvature. For cases where heat or mass transfer through the liquid should be considered, internal circulation, through the convective and advective aspects, can have major impact on diffusion time scales. At lower $Re$ values where separation does not occur, near symmetry in the $x$-direction for both the pressure profile and the interface shape is expected. The literature has much discussion on Hill's spherical vortex, where the vorticity is proportional the cylindrical radial distance thoughout the full sphere with zero vorticity outside the sphere. The fluid in the sphere remains there, while a flow occurs around the outside of the sphere. Both the external and internal flows are axisymmetric.  Some extensions of the Hill spherical vortex account for azimuthal swirl, which is not relevant to our present analysis. See \cite{Hill}, \cite{Lamb},  \cite{Saffman}, and \cite{O’Brien1961}  for general background discussion on the spherical vortex.  There are interesting studies by  \cite{Fraenkel1970}, \cite{Fraenkel1972} and \cite{Norbury1972,Norbury1973} examining axisymmetric vortex rings with symmetry in the $x$-direction that present three domains of flow: an internal vortex ring, a circulating irrotational flow surrounding the vortex ring but remaining within a closed boundary, and an external irrotational flow around the closed boundary. Here, the cross-sections of the flow in a cylindrical $r, x$ plane do not show circular shapes  for the vortex ring or encapsulated fluid boundaries.  For our droplet application, liquid-phase vorticity is generated by shear at the gas interface. This character creates a situation where vorticity should exist adjacent to the interface; so, models where the liquid-phase vortex ring does not contact the gas interface are not useful for a droplet study.    Only in the Hill's vortex limit where the vortex ring fills the encapsulated flow region do we have symmetries in both $r$ and $x$.
\cite{O’Brien1961} briefly discusses extension from a sphere to an ellipsoid which is very relevant to our interests. Similar behavior to the spherical vortex can be found.

Consider an ellipsoidal droplet with axisymmetric internal incompressible, inviscid, steady flow. Here, the steady-state will be used to approximate a quasi-steady situation where the internal liquid flow responds quickly to any change in relative gas-droplet velocity. Vorticity fills the volume of the droplet with vorticity magnitude $\omega$ proportional to cylindrical radial position $\omega = k r$ where the constant $k$ will be determined below to match other parameters. The stream function $\Psi_l$ inside the ellipsoid is given as $\Psi_l = Ur^2[ (r/a)^2 + (x/b)^2 -1]/2$. 
The streamlines $\Psi_l = 0$ are forced here along the axis $r=0$ and the ellipsoid surface $(r/a)^2 + (x/b)^2 = 1$  with $a$ and $b$ as the semi-major axis and semi-minor axis, respectively. The value of the characteristic velocity $U$ will be taken as the $U_{l,max}$ value given by Equation (\ref{maxvelocity}).

The axial velocity component $u$, radial velocity component $v$, and the vorticity $\omega$ can be written as functions of the derivatives of $\Psi_l$.
\begin{eqnarray}
\Psi_l = Ur^2[ (r/a)^2 + (x/b)^2 -1]/2  \;\; ; \; \;u =\frac{1}{r}\frac{\partial \Psi_l}{\partial r}  \;\; ; \;\;\; v = -\frac{1}{r}\frac{\partial \Psi_l}{\partial x} \;\;\; ; \;\;\; \omega  =  
\frac{1}{r}\frac{\partial^2 \Psi_l}{\partial x^2} +
\frac{1}{r}\frac{\partial^2 \Psi_l}{\partial r^2}   - \frac{1}{r^2}\frac{\partial \Psi_l}{\partial r} 
\label{liquidstream}
\end{eqnarray}
For the prescribed stream function, we have
\begin{eqnarray}
u = U\Big[ 2\Big(\frac{r}{a}\Big)^2 +  \Big(\frac{x}{b}\Big)^2 -1\Big] \;\;\; ; \;\;\; v = -U \frac{rx}{b^2} \;\;\; ; \;\;\; \omega = -\frac{U}{a}\Big[  4 + \Big(\frac{a}{b}\Big)^2  \Big]\frac{r}{a}
\label{vorticity}
\end{eqnarray}
This equation prescribes the above mentioned constant to have $k = - (U/a^2)[4 +(a/b)^2]$. We see therefore clockwise rotation for positive $r$ values. The characteristic velocity $U$ is the largest value of the velocity inside the ellipsoid. It is achieved at $x=0, r=a$ where $u=U, v=0$ and  at $x=0, r=0$ where $u=-U, v=0$.  Both velocity components become zero at $x=0, r= a/\sqrt{2}$; this point is in the center  of the vortex ring and denotes the circular ring that is central to the re-circulating liquid.  The magnitude of the tangential velocity $U_s$ along the surface of the ellipsoid is given as 
\begin{eqnarray}
\frac{U_s}{U} &=& \frac{\sqrt{u_s^2 +v_s^2}}{U} =\frac{r}{a}\Big[1 + \Big(\big(\frac{a}{b}\big)^2 -1  \Big) \Big(\frac{x}{b}\Big)^2  \Big]^{1/2} \nonumber \\ &=&     \Big[ 1- \Big(\frac{x}{b}\Big)^2\Big]^{1/2}\Big[1 + \Big(\big(\frac{a}{b}\big)^2 -1  \Big) \Big(\frac{x}{b}\Big)^2  \Big]^{1/2}
\label{surfacespeed}
\end{eqnarray}
For a sphere, the non-dimensional surface velocity is simply $r/a$.

The chosen stream function must and does satisfy the equations for conservation of mass and momentum for a steady, incompressible, inviscid fluid. The equation of continuity gives
\begin{eqnarray}
    \frac{\partial (ru)}{\partial x} + \frac{\partial (rv)}{\partial r} = \frac{\partial^2 \Psi_l}{\partial x \partial r} - \frac {\partial^2 \Psi_l}{\partial r \partial x} =0
\end{eqnarray}
For momentum, we shall use the vorticity equation where the vector $\vec{\omega} =\omega \vec{e}_{\phi} = kr\vec{e}_{\phi}$ where $\vec{e}_{\phi}$ is the unit vector in the azimuthal $\phi$ direction.
\begin{eqnarray}
u\frac{\partial \omega}{\partial x} + v\frac{\partial \omega}{\partial r} &=& \Big(\frac{\vec{\omega}}{r}\Big)\cdot \frac{\partial (v \vec{e}_{r})}{\partial \phi} \nonumber \\
u\frac{\partial (kr)}{\partial x} + v\frac{\partial (kr)}{\partial r} &=& \Big(\frac{v\vec{\omega}}{r}\Big)\cdot\frac{\partial \vec{e}_r}{\partial \phi} \nonumber \\  0 + kv &=&\frac{v}{r}\vec{\omega}\cdot \vec{e}_{\phi}
= v\frac{\omega}{r} = k v
\end{eqnarray}
It is known that $\omega = F(\Psi_l)r$ will satisfy this vorticity conservation equation. In our case, the function $F$ is a simple constant. Once the velocity field is known, the pressure through the circulating liquid is determined by Bernoulli's equation. 
\begin{eqnarray}
p_{l}  &=& p_{l, stag} - \frac{\rho_{l}}{2} (u^2 + v^2) \nonumber \\  &=& p_{l, stag} - \frac{\rho_{l}U^2}{2} \Bigg[\Big[ 2\Big(\frac{r}{a}\Big)^2 +  \Big(\frac{x}{b}\Big)^2 -1 \Big]^2 + \frac{r^2x^2}{b^4} \Bigg]
\label{newpressure}
\end{eqnarray}

The $u$ velocity component is symmetric in $x$ while the $v$ component is antisymmetric in $x$ and goes to zero value along the $r =0$ axis.
Both velocity components become zero at $x = 0 , r =a/\sqrt{2}$. Thus, with sufficiently rapid circulation, the convective impact is to reduce the character length for diffusion several-fold and the characteristic diffusion time by an order of magnitude. This effect has been well noted for spherical droplets by \cite{Prakash1978}, \cite{Prakash1980}, \cite{Sirignano1983},  \cite{Abramzon}, and \cite{Sirignano_2010},

This internal-circulation analysis could be matched to the surface-curvature analysis by setting the semi-major axis $a = 1 +5 \delta/12$ and the semi-minor axis $ b = 1- \delta/2$. Of course, as $\delta \rightarrow 0$, Hill's spherical vortex is obtained. 
\begin{figure}[thbp]
 \subfigure [Stream function $\Psi_l$ for $We =0$. ]{
  \includegraphics[height = 4.6cm]{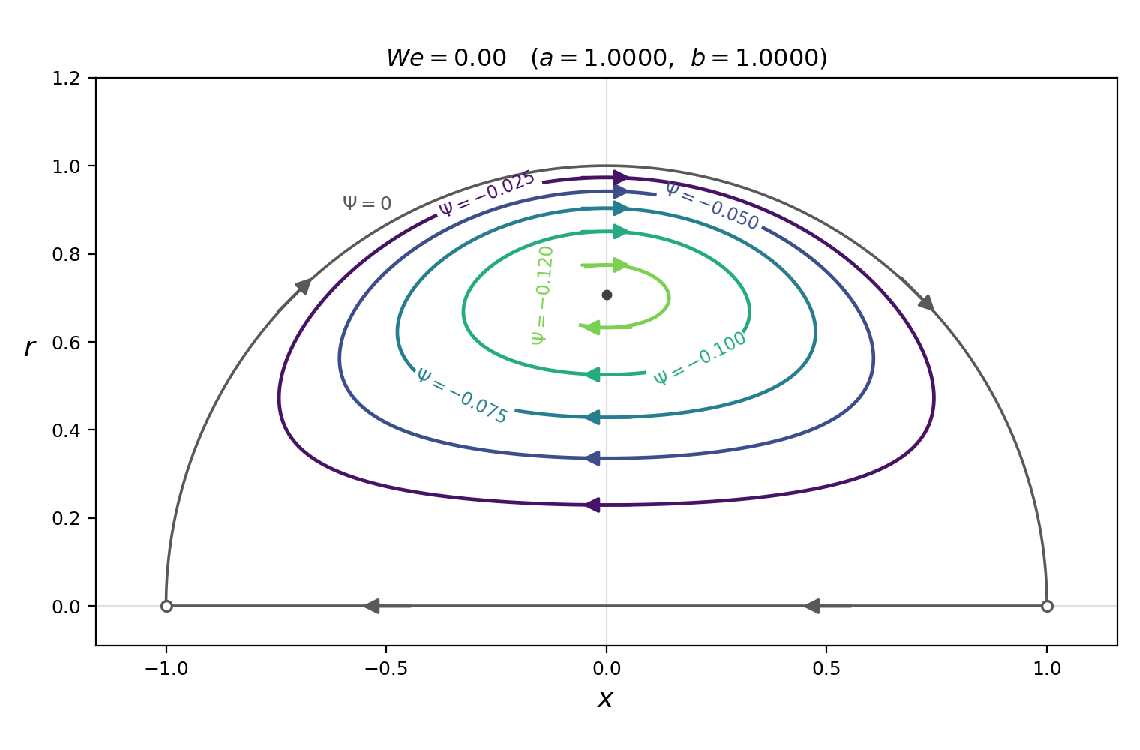}}
  \subfigure [Stream function $\Psi_l$ for $We =0.25$. ]{
  \includegraphics[height = 4.6cm]{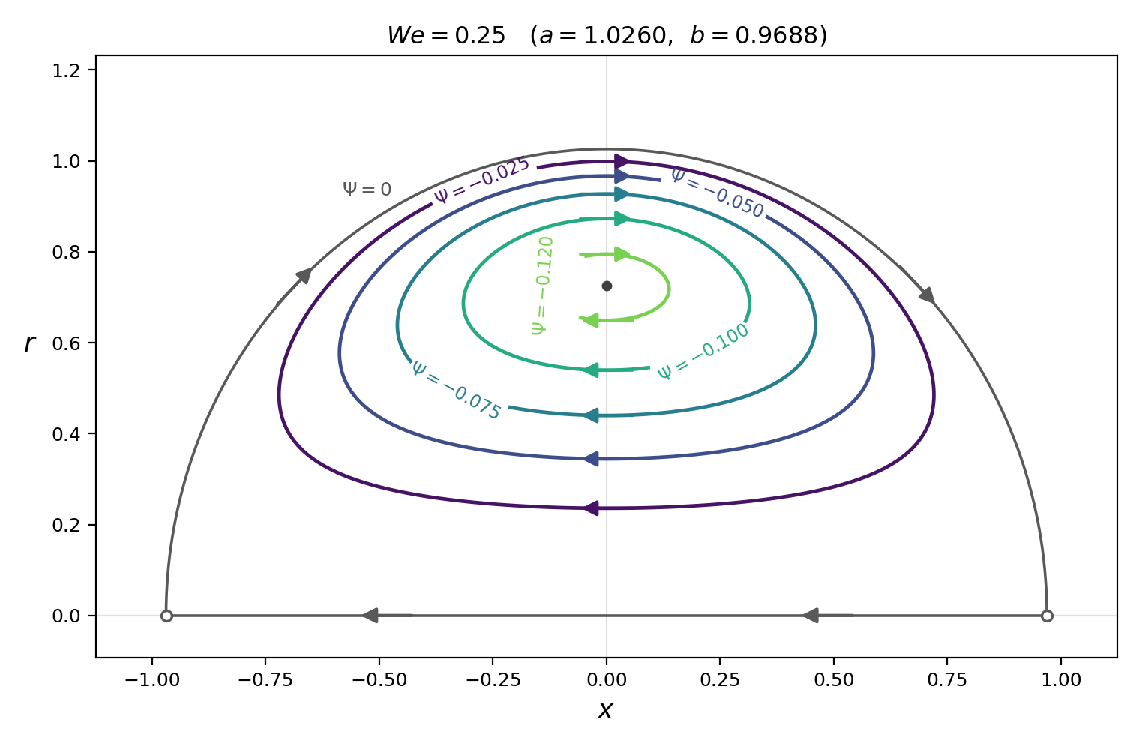}}
\\
   \subfigure[Stream function $\Psi_l$ for $We =0.50$.]{
  \includegraphics[height = 4.6cm]{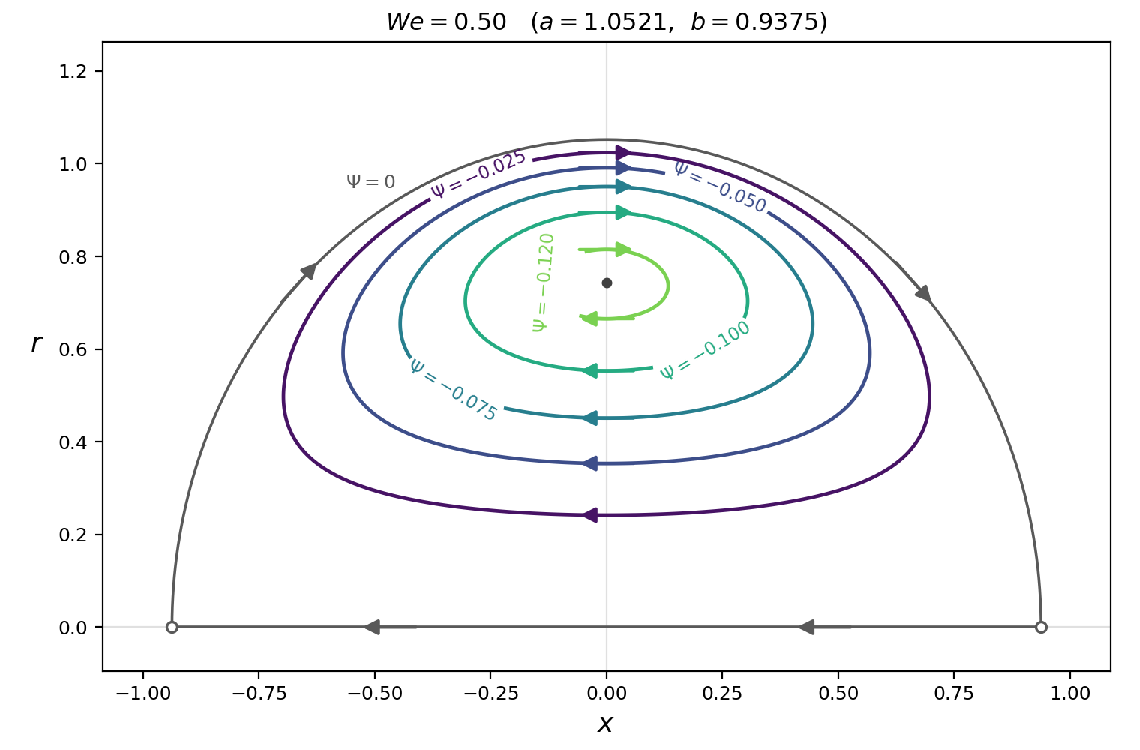}}
   \subfigure[Stream function $\Psi_l$ for $We =0.75$.]{
  \includegraphics[height = 4.6cm]{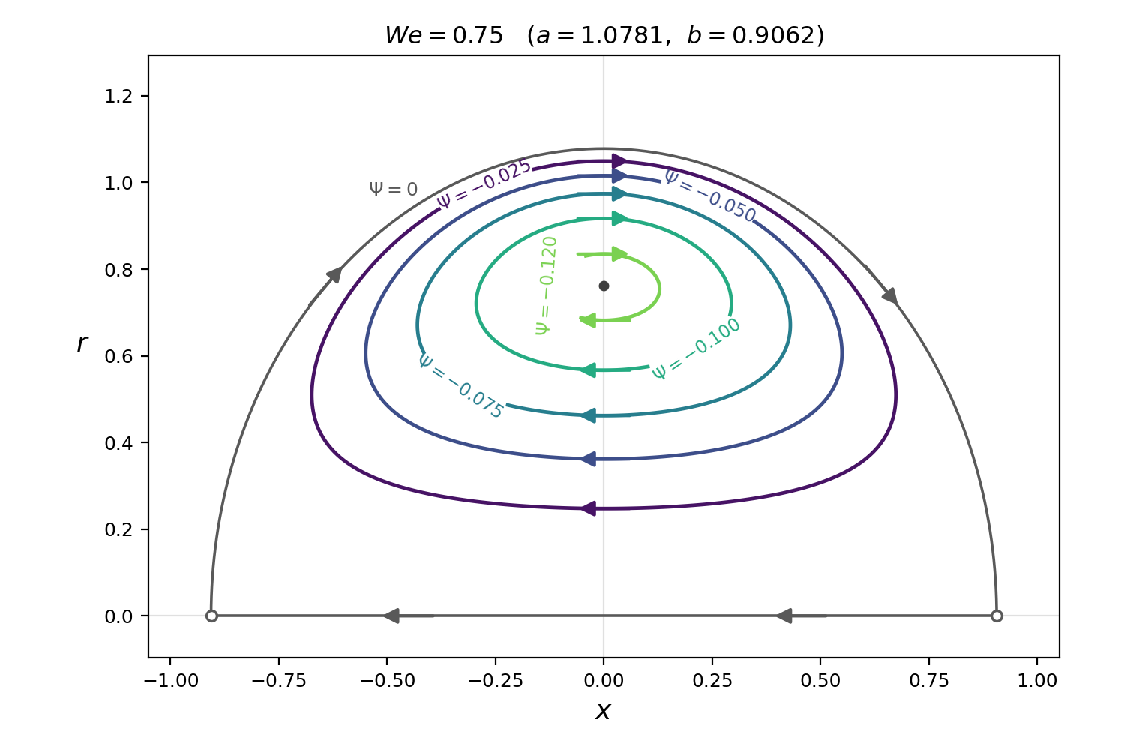}}
  \\
   \subfigure[Stream function $\Psi_l$ for $We =1.00$.]{
  \includegraphics[height = 4.6cm]{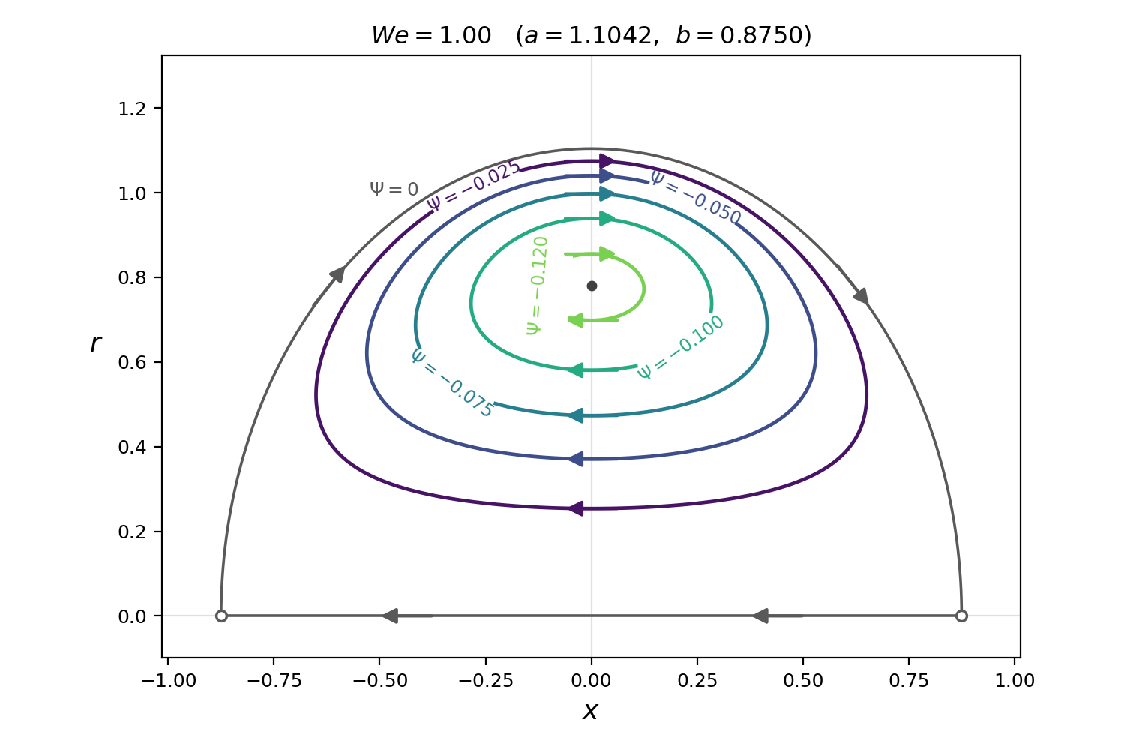}}
   \subfigure[Surface Shape]{
  \includegraphics[height = 4.6cm]{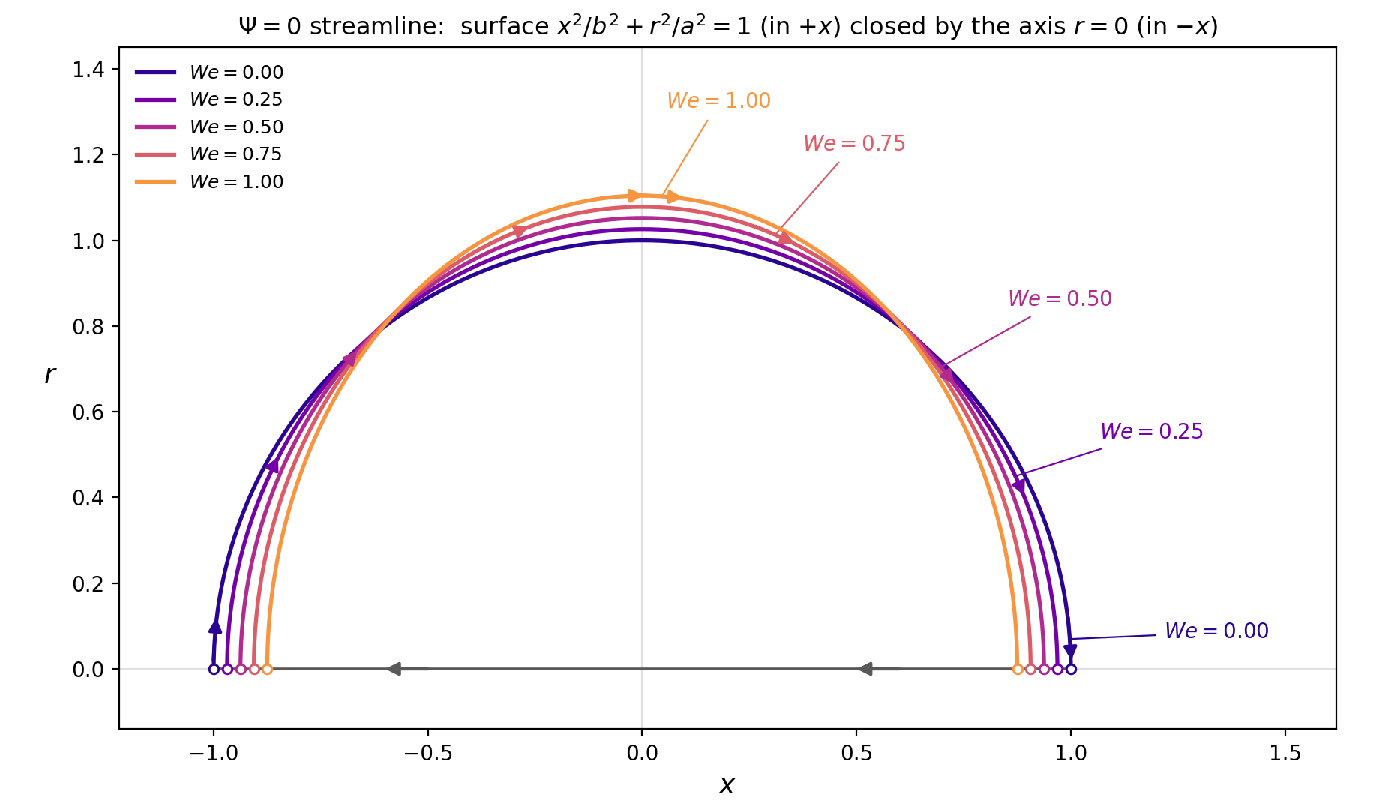}}
   \caption{Stream function contours  and droplet shape from Equations (\ref{liquidstream}, \ref{vorticity}) at selected Weber Number $We$ values, $0\leq We \leq 1$. $x$ is normalized by the  spherical radius for the $We =0$ case.}
  \label{Stream}
  \end{figure}
Figure \ref{Stream} gives the contours for selected values of 
nondimensional stream function values, now normalized by $U R_0^2$ and Weber number $We$. $We =0$ is the spherical droplet case. $x, r, a,$ and $b$ are all normalized by the radius $R_0$ of the spherical droplet with the given volume. The semi-major axis grows while the semi-minor axis decreases with increasing $We$. The ratio of the vortex-ring-center height (or circle radius) to the semi-major axis remains constant at $1/\sqrt{2}$ as $We$ varies. 
The magnitude of the stream function peaks at that center ring at the same value of $\Psi_l = -0.125$ for all $We$ values.
All velocity values scale with $U$ the maximum liquid velocity which occurs at $r= a, x=0$. Until we match the liquid and gas flows, we cannot determine $U(We)$.\\

\begin{figure}[thbp]
 \subfigure [Normalized surface velocity $U_s/U$. ]{
  \includegraphics[height = 4.6cm]{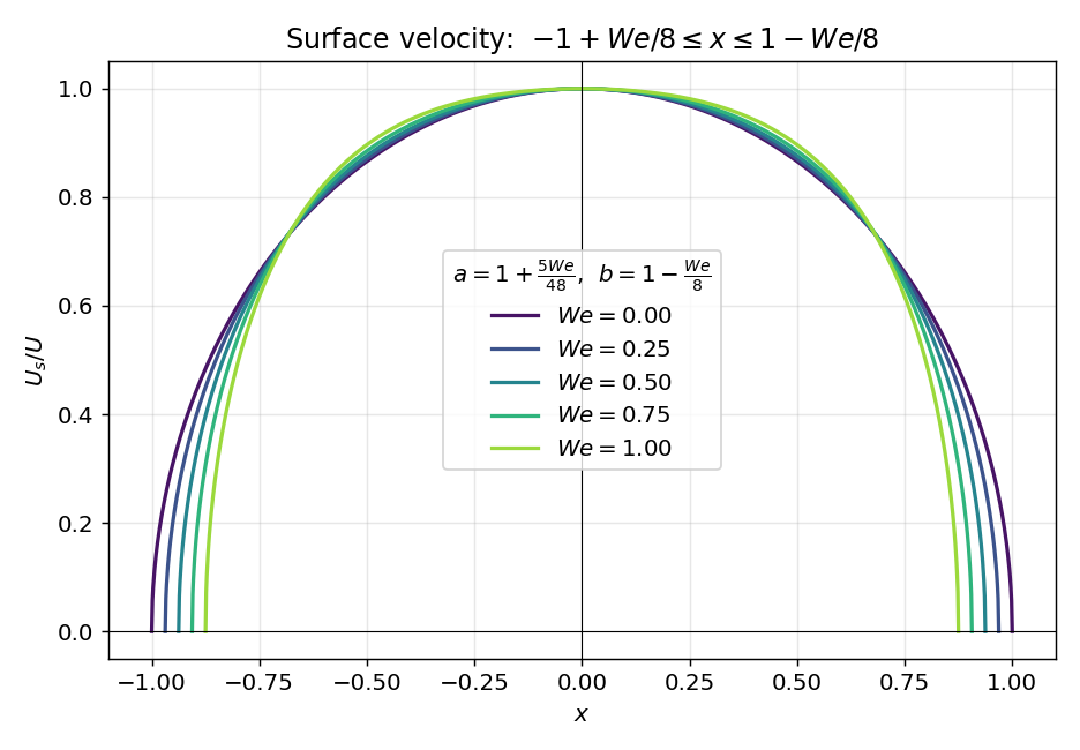}}
  \subfigure [Normalized surface pressure variation $(p_{l,stag} - p_l)/(\rho_l U^2)$ . ]{
  \includegraphics[height = 4.6cm]{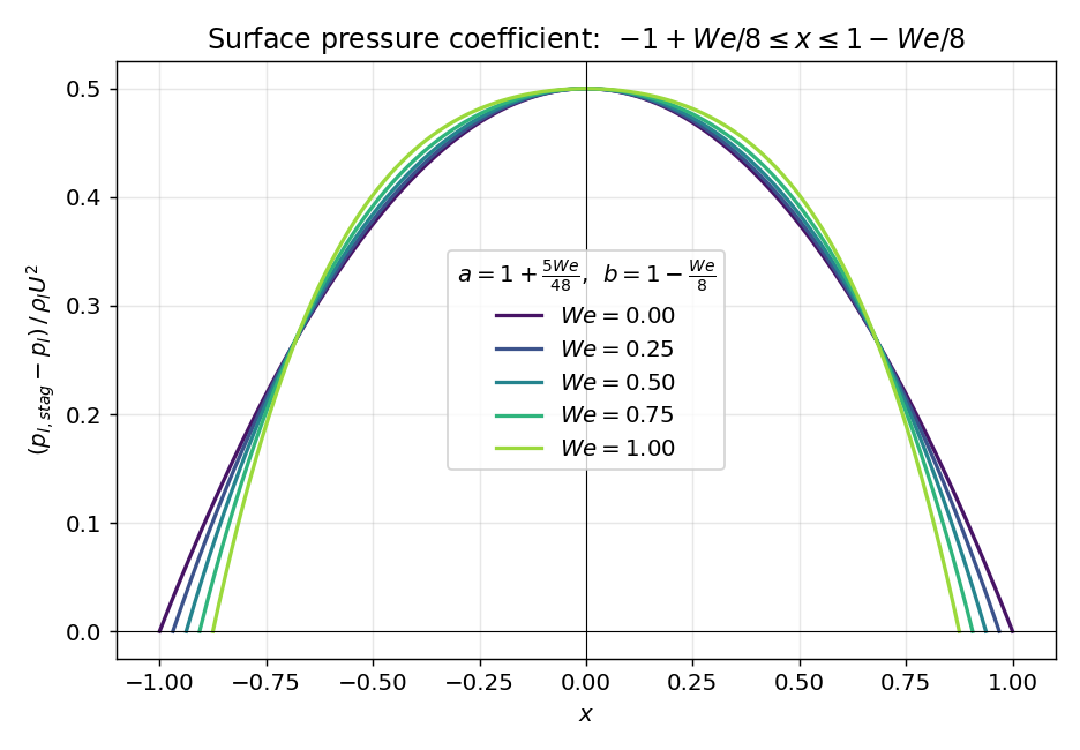}}
   \caption{Surface pressure and surface velocity variation from Equations (\ref{surfacespeed}, \ref{newpressure}) at selected Weber Number $We$ values, $0\leq We \leq 1$. $x$ is normalized by the  spherical radius for the $We =0$ case. }
  \label{Surface}
  \end{figure}

Figure \ref{Surface} shows the variations of surface velocity and surface pressure drop for the five different chosen $We$ values. The normalization is with the peak values at $x=0$. Until we have a gas-phase analysis for matching, the value of the peak surface pressure difference or surface velocity cannot be prescribed. 
  \\
Although the path will not be pursued here, the pressure along the surface given by Equation (\ref{newpressure}) can be useful in providing a higher-order correction to the surface curvature. This requires combination  with a correction of the gas-phase surface velocity and pressure due to the perturbed droplet shape.

\section{Gas-phase Potential Flow over Oblate Spheroid} \label{potent}

The gas flow will be described as an axisymmetric potential flow with a uniform free stream velocity $U_{\infty}$ in the positive $x$-direction. A thin surface boundary layer will be evaluated later. The $Re$ is low enough to avoid separation. The gas velocity potential $\Phi$ is created as the sum of the free-stream value $U_{\infty}x$ and a doublet image in the droplet. The spherical droplet can be described using a point doublet with $x$-orientation at the center of the sphere. We will examine a distributed doublet in the $x=0$ plane. A general discussion of a doublet  distributed over a general planar surface is given by \cite{KatzPlotkin}. We will follow that general concept to have a ring doublet which has strength on a circular plate of zero thickness (in the $x$-direction), lies in the $x=0$ plane with zero-thickness in the $x$-direction, and is centered on the origin at $x=0, r=0$ in the axisymmetric field. It is annular and extends from $r = \sigma$ to $r= \sigma + \Delta \sigma$.  At each point on this annular plate, the doublet can be viewed as the limit of bringing together at $x=0$ a source from the upstream side $x<0$ and a sink from the downstream side $x>0$. The ring doublet is an image for the gas flow external to the droplet and does not predict flow within droplet. The author is unaware of prior use of a ring doublet. We will also add a weaker point-doublet at the origin to make small adjustment in the surface boundary to match the above portions of our analysis. 

\begin{figure}[thbp]
\centering
  \includegraphics[height = 4.6cm]{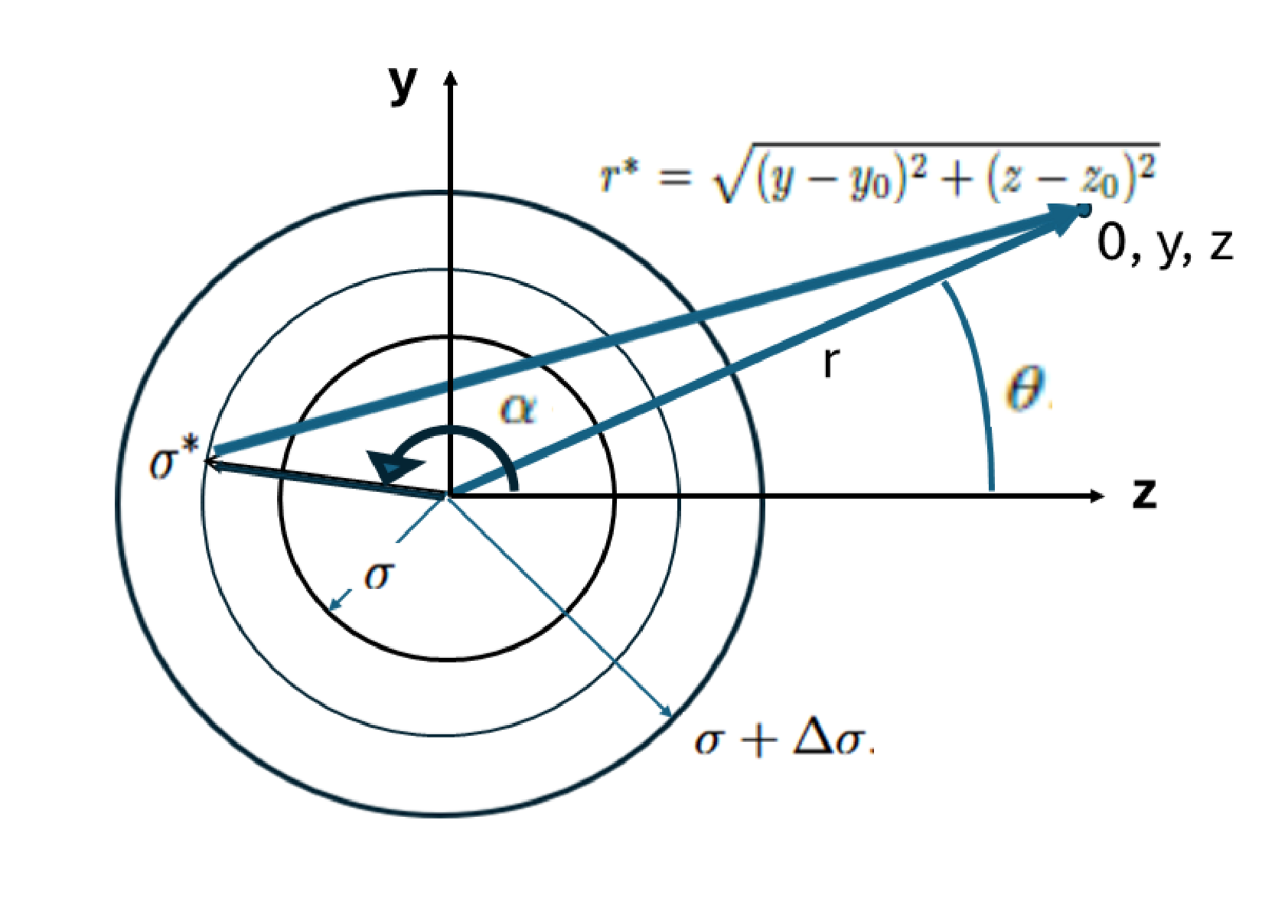}
  \caption{Scheme for ring doublet plate lying in the $x=0$ plane. Radial position $\sigma^*$, of point on doublet plate, radial position of impacted flow point $r$, distance $r^*$ between these two points, and the ring plate inner and outer boundaries $\sigma$ and $\sigma + \Delta \sigma$ are shown.}
  \label{Doublet}
  \end{figure}
Let us start with a Cartesian coordinate system and later return to cylindrical coordinates. The $x=0$ plane is a $y-z$ plane here with $r =\sqrt{y^2 + z^2}$. Figure \ref{Doublet} shows the annular doublet ring scheme in the $x=0$ plane where it lies. Of course, it impacts through all three dimensions. The doublet ring of strength $\mu^*$ has $\sigma^{*2} = y_o^2 + z_0^2$ describing the radius for a point of the doublet annular plate, with $\sigma \leq \sigma^* \leq \sigma +\Delta \sigma$.  So, $y_0$ and $z_0$ vary around the ring. 
The resulting velocity potential $\Phi(x,y,z)$ from the ring alone is
\begin{eqnarray}
    \Phi = -\frac{x}{4 \pi}\int\int \mu^* \frac{dy_0dz_0}{\big[x^2 + (y-y_0)^2 +(z-z_0)^2\big]^{3/2}}
    \label{Phi}
\end{eqnarray}
We start with a surface integral but will soon create a delta function strength on the thin ring.
In the $x=0$ plane $r^* = \sqrt{(y-y_0)^2 +(z-z_0)^2}$ is the distance from a point $0, y_0, z_0$ on  the doublet to a general $0, y, z$ point in the plane. $\sqrt{x^2 + (y-y_0)^2 +(z-z_0)^2}$ would be the distance to an $x, y, z$ point outside of the plane. In the plane, we define $\theta = \arctan (y/z)$  and  $\alpha = \arctan (y_0/z_0)$. The law of cosines yields $r^{*2}= r^2 + \sigma^{*2} -2 r \sigma^* \cos(\alpha - \theta)$. Define $\alpha^* \equiv \alpha - \theta$. The substitution into Equation (\ref{Phi}) yields
\begin{eqnarray}
    \Phi &=& -\frac{x}{4 \pi}\int\int \mu^*\frac{dy_0dz_0}{\big[x^2 + r^2 + \sigma^{*2} -2 r \sigma^* \cos(\alpha - \theta)\big]^{3/2}}
  \nonumber \\
    &=& -\frac{ x}{4 \pi}\int_{\sigma}^{\sigma + \Delta \sigma}\int_0^{2\pi}\mu^*\frac{\sigma^* d\alpha^*d\sigma^* }{\big[x^2 + r^2 + \sigma^{*2} -2 r \sigma^* \cos(\alpha^*)\big]^{3/2}}
        \label{Phi2}
\end{eqnarray}
Let $\Delta \sigma^* \rightarrow 0$ and $\mu^* \rightarrow \infty$ ( using a Dirac delta function ) while keeping the finite value for $\mu = \mu^*\Delta \sigma$, thereby forming a ring of zero thickness.  Subsequently,
\begin{eqnarray}
    \Phi = -\frac{ \mu \sigma x}{4 \pi}\int_0^{2\pi}\frac{d\alpha^*}{\big[x^2 + r^2 + \sigma^2 -2 r \sigma \cos(\alpha^*)\big]^{3/2}}
        \label{Phi3}
\end{eqnarray}

We proceed by using $R_0$, the radius of the sphere with equivalent volume to our spheroid, to normalize lengths and the free-stream velocity $U_{\infty}$ to normalize velocities. For our low $We$ interest, we have $\varepsilon \equiv \sigma/R_0 <<1$ and can use it as a perturbation parameter. Simultaneously, we increase the $\mu$ value as $\sigma$ decreases, so that $-\mu\sigma = U_{\infty}R_0^3$, a constant. The following integrals will be useful to us.
\begin{eqnarray}
I_1 \equiv \frac{1}{4 \pi} \int_0^{2\pi} \frac{d\alpha^*}{\Big[ \frac{r^2 +x^2}{R_0^2} + \varepsilon^2 - 2\varepsilon \frac{r}{R_0} \cos \alpha^*  \Big]^{3/2}}
\nonumber \\
I_2 \equiv \frac{1}{4 \pi} \int_0^{2\pi} \frac{d\alpha^*}{\Big[ \frac{r^2 +x^2}{R_0^2} + \varepsilon^2 - 2\varepsilon \frac{r}{R_0} \cos \alpha^*  \Big]^{5/2}}
\nonumber \\
I_3 \equiv\frac{1}{4 \pi}  \int_0^{2\pi} \frac{\cos \alpha^* \;d\alpha^*}{\Big[ \frac{r^2 +x^2}{R_0^2} + \varepsilon^2 - 2\varepsilon \frac{r}{R_0} \cos \alpha^*  \Big]^{5/2}}
\label{I}
\end{eqnarray}

Let us add the free-stream potential $U_{\infty}x$ to the doublet ring. We also add a weak point-doublet of strength $U_{\infty} \varepsilon^2/2$ at the origin that will be helpful in creating a spheroidal interface. Now, we have the potential function and the velocity components
\begin{eqnarray}
\Phi &=& U_{\infty}\Big[x + x I_1 +\frac{\varepsilon^2 x R_0^3}{4(x^2 +r^2)^{3/2}}\Big]
\nonumber \\
u &=& \frac{\partial \Phi}{\partial x} = U_{\infty}\Big[ 1  
+ I_1 - \frac{3x^2 I_2}{ R_0^2}  + \frac{\varepsilon^2 (r^2 - 2 x^2) R_0^3}{4(x^2 +r^2)^{5/2}}  \Big] 
\nonumber \\
v &=& \frac{\partial \Phi}{\partial r} = -3 U_{\infty}\Big[\frac{xr}{R_0^2} I_2
- \frac{\varepsilon x}{R_0}I_3  - \frac{\varepsilon^2 xr R_0^3}{4(x^2 +r^2)^{5/2}} \Big]
\label{Phi4}
\end{eqnarray}
It can be shown that $\Phi$ satisfies Laplace's equation. In particular, the contribution to the flow from any point on the doublet ring plate satisfies the equation, as do the contributions from the free stream and the weak doublet at the origin.  It is simple to show this conclusion using Equation (\ref{Phi}) in Cartesian coordinates or using Equation (\ref{Phi3}) in cylindrical coordinates. In the latter case, it must be realized that $\alpha^*$ depends linearly on the coordinate $\theta$. The satisfaction of Laplace's Equation can shown for each individual doublet by differential analysis at each point on the doublet ring and inside the double integral over the ring; the linear character assures that Laplace's result will apply after the integration over the surface.  In the limit $\varepsilon \rightarrow 0$, $I \rightarrow 0$, while $I_1$ and $I_2$ will give $\Phi, u$, and $v$ the form for an axisymmetric combination of a free stream and a doublet at the origin, namely the flow around a sphere.  Specifically, we obtain for that limiting flow $I_1 = 1/(2[(x^2 + r^2)/R_0^2]^{3/2}), I_2 = 1/(2[(x^2 + r^2)/R_0^2]^{5/2}),$ and $I_3 =0$.  In that special case, it follows that, along any spherical surface centered on the origin (including the spherical droplet surface), $\Phi$ is linear in $x$, $u$ behaves as a constant plus a term proportional to $x^2$, and $v$ is the sum of two terms, one proportional to $x$ and the other proportional to $x^3$.

A stream function $\Psi(r, x)$ can exist only in two spacial variables. So, it cannot be constructed following the path for the velocity potential creation, which started in three dimensions and by integration over $\alpha^*$ was stated an an axisymmetric function. 
However, we may take the results in Equation (\ref{Phi4}) and set $ \partial \Psi/ \partial r = ur$ and $\partial \Psi/ \partial r = -vr$. We define a new definite integral $I_4$ and use the condition that a certain combination of definite integrals collectively has a cyclic integrand and produces zero upon integration. We integrate $\partial \Psi/ \partial x $ first for the ring-doublet portion alone to obtain a solution with an added unknown function of $r$. Then, we show the derivative with respect to $r$ is satisfied with the added function set to zero.
\begin{eqnarray}
 \frac{1}{U_{\infty}R_0} \frac{\partial \Psi_{ringdoublet}}{\partial r} &=& \frac{r}{R_0} I_1 - 3\frac{rx^2}{R_0^3} I_2  \;\; ; \;\;
 \nonumber \\
 \frac{1}{U_{\infty}R_0}\frac{\partial \Psi_{ringdoublet}}{\partial x} &=& 3\Big[\frac{xr^2}{R_0^3} I_2  -\varepsilon \frac{xr}{R_0^2} I_3  \Big]
\nonumber \\  
I_4 &\equiv& \frac{1}{4\pi} \int_0^{2\pi} \frac{\cos \alpha^* d\alpha^*} {\Big[ \frac{r^2 +x^2}{R_0^2} + \varepsilon^2 - 2\varepsilon \frac{r}{R_0} \cos \alpha^*  \Big]^{3/2}}  \nonumber \\
\frac{1}{4 \pi}\int_0^{2 \pi} \frac{\partial}{\partial \alpha^*} \frac{\sin \alpha^*}{\Big[ \frac{r^2 +x^2}{R_0^2} + \varepsilon^2 - 2\varepsilon \frac{r}{R_0} \cos \alpha^*  \Big]^{3/2}} d\alpha^* &=&
\frac{1}{4\pi}\int_0^{2\pi}\Bigg[ \frac{\cos \alpha^*} {\Big[ \frac{r^2 +x^2}{R_0^2} + \varepsilon^2 - 2\varepsilon \frac{r}{R_0} \cos \alpha^*  \Big]^{3/2}} \nonumber \\
&&-\frac{3r \varepsilon \sin^2 \alpha^*}{\Big[ \frac{r^2 +x^2}{R_0^2} + \varepsilon^2 - 2\varepsilon \frac{r}{R_0} \cos \alpha^*  \Big]^{5/2}} \Bigg]d\alpha* =0
\nonumber \\
   \frac{\Psi_{ringdoublet}}{U_{\infty}R_0^2} =&&-\Big(\frac{r}{R_0}\Big)^2 I_1 + \frac{\varepsilon r}{R_0}I_4  \nonumber \\
   \frac{\Psi}{U_\infty R_0^2} =&&\frac{r^2}{2R_0^2}-\Big(\frac{r}{R_0})^2 I_1 + \frac{\varepsilon r}{R_0}I_4  -\frac{\varepsilon^2r^2R_0}{4(x^2 +r^2)^{3/2}}
\end{eqnarray}
In the Appendix, we show the identify between this ring doublet and the confluence of two concentric ring vortices with opposing directions of rotation.
Later, using the perturbation method, we will approximate the stream function. Using the exact solution, we can and will plot the streamline curves for the axisymmetric flow.

The two stagnation points can be found by setting $r=0$ in the three integrals and seeking the point where both $u(0,x) =0$ and $v(0,x) =0$. (Of course, $v=0$ at $r=0$ for any $x$ value.) For $r=0$, the three integrals in Equation (\ref{I}) are easily integrated. $I_1 =1/(2[x^2/R_0^2 +\varepsilon^2]^{3/2}); I_2  =1/(2[x^2/R_0^2 +\varepsilon^2]^{5/2})$, and $I_3 =0$ along $r=0$. 

Then, using Equation (\ref{Phi4}) with $r=0$, the stagnation points $X^* = x^*/R_0 \equiv \pm \sqrt{1-\beta\varepsilon^2}$ are determined. Those points are given by $X^2 = 1 - \beta\varepsilon^2$ where $\beta$ is the solution of the algebraic relation 
\begin{eqnarray}[1 +(1-\beta)\varepsilon^2]^{5/2}\Big[1 -\frac{\varepsilon^2}{2(1 -\beta \varepsilon^2)^{3/2}}\Big] -1 +(\beta +1/2)\varepsilon^2= 0.
\end{eqnarray}
A series expansion of the algebraic relation in $\varepsilon^2$ shows that  $ \beta = 5/3 +O(\varepsilon^2)$ ,  and the stagnation points occur at  $r/R_0= 0, X^2 =1 -(5/3)\varepsilon^2 +O(\varepsilon^4)$ or thereby at both $r/R_0=0, X = 1- (5/6)\varepsilon^2 +O(\varepsilon^4)$ and $r/R_0=0, X = -1 + (5/6)\varepsilon^2 +O(\varepsilon^4)$. Thus, the semi-minor axis which equals the particular $x$ (or $X$) value is decreased from the value for a sphere of the same volume. Of course, the exact value for the semi-minor axis and these stagnation point locations can be found without expansion  by numerical iteration using the above exact algebraic relation.

\begin{figure}[thbp]
 \subfigure [Droplet shape $r(X)$ . ]{
  \includegraphics[height = 4.6cm]{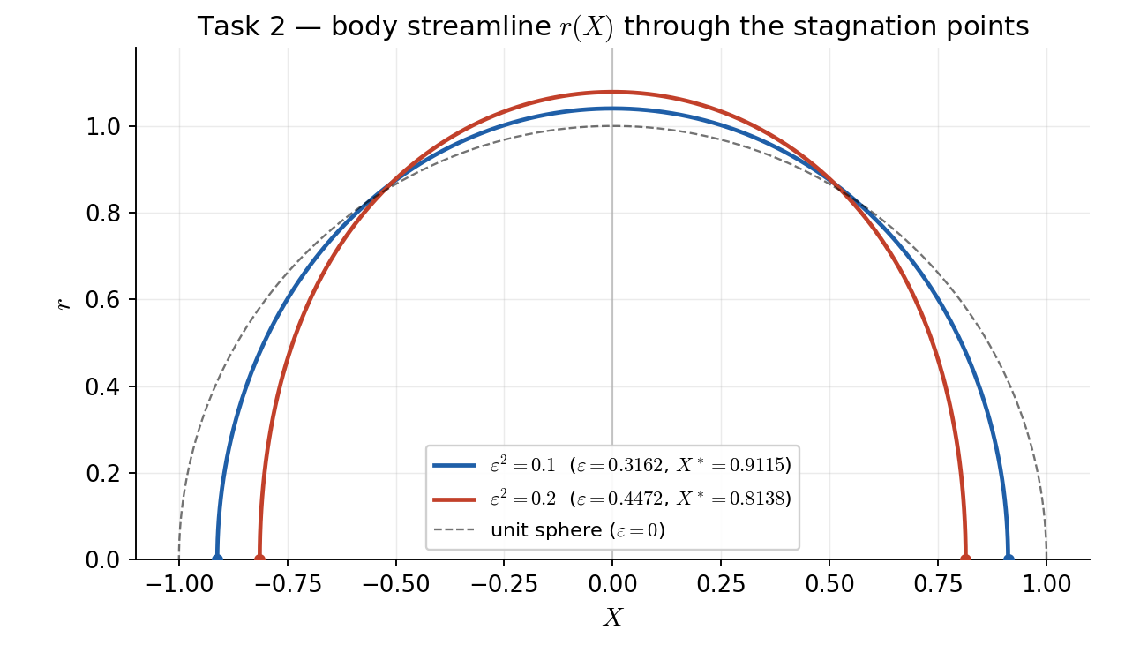}}
   \subfigure [Ellipsoid comparison. $r^2/R_0^2$ . ]{
  \includegraphics[height = 4.6cm]{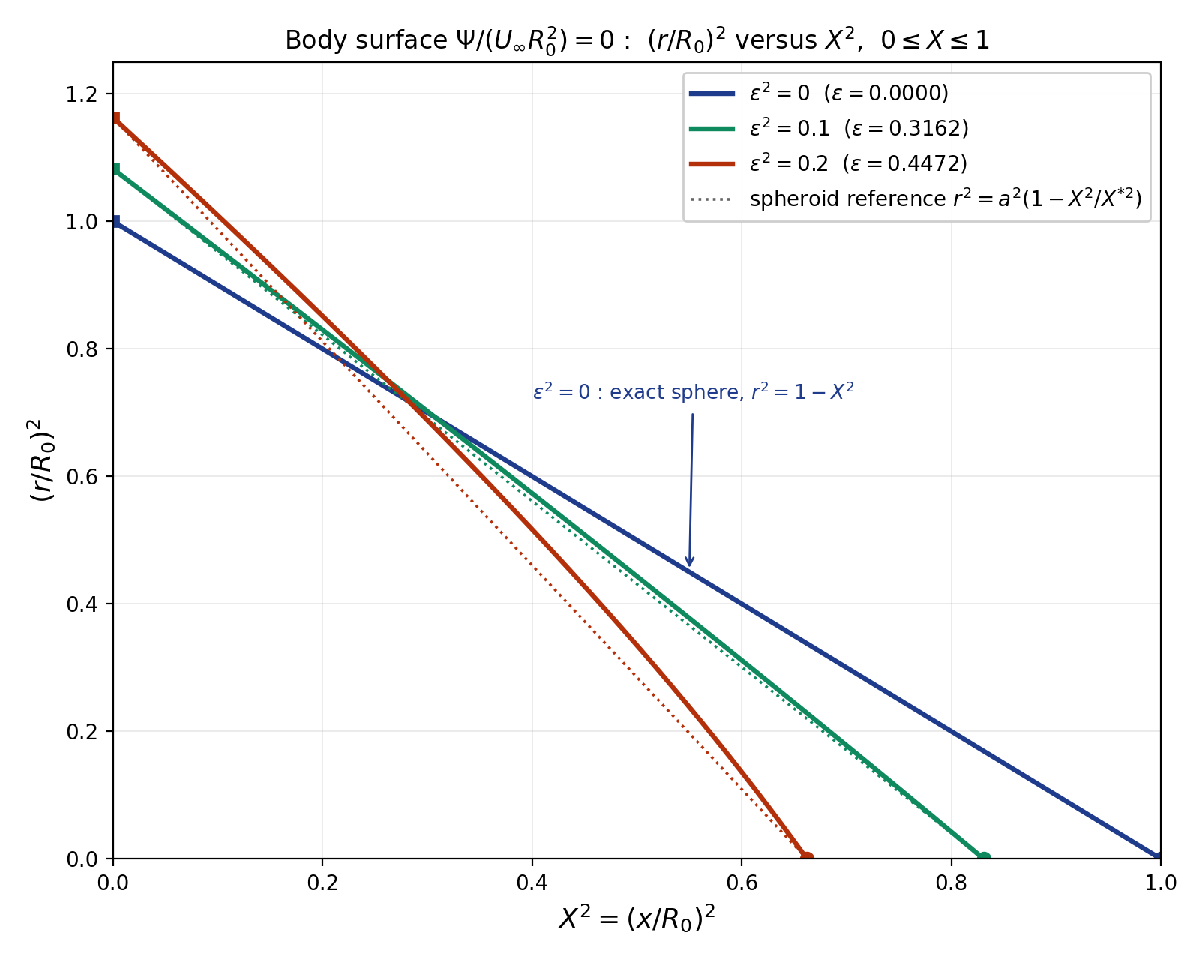}}
  \\
  \subfigure [Gas velocity potential function $\Phi$ at surface. ]{
  \includegraphics[height = 4.6cm]{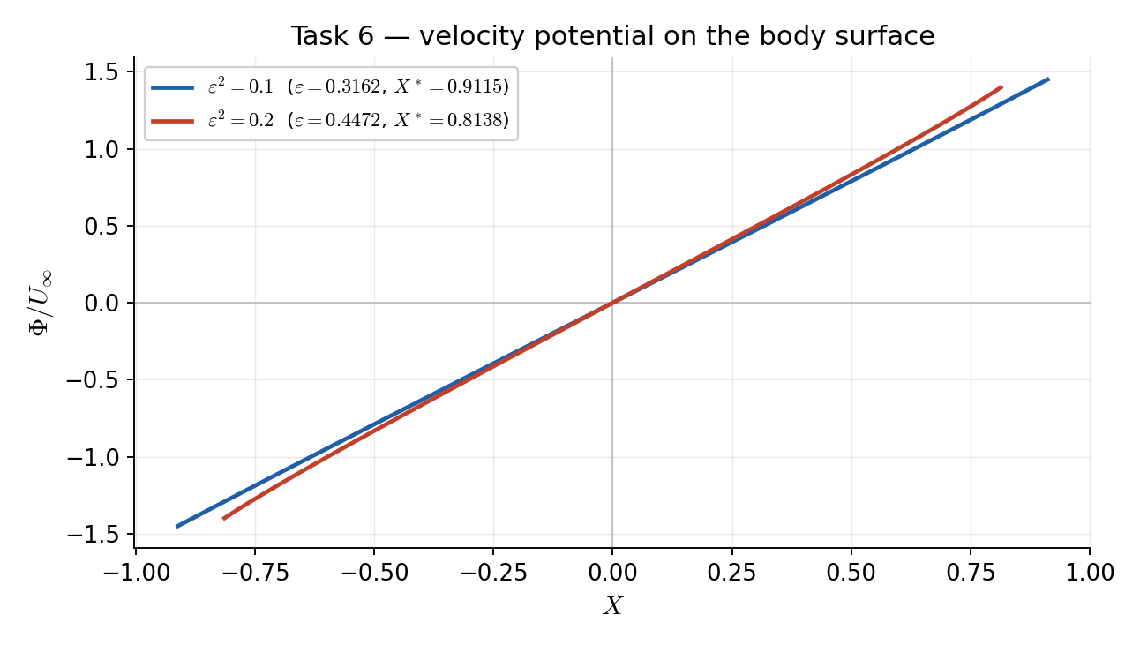}}
   \subfigure[Axial gas velocity component $u$ at surface.]{
  \includegraphics[height = 4.6cm]{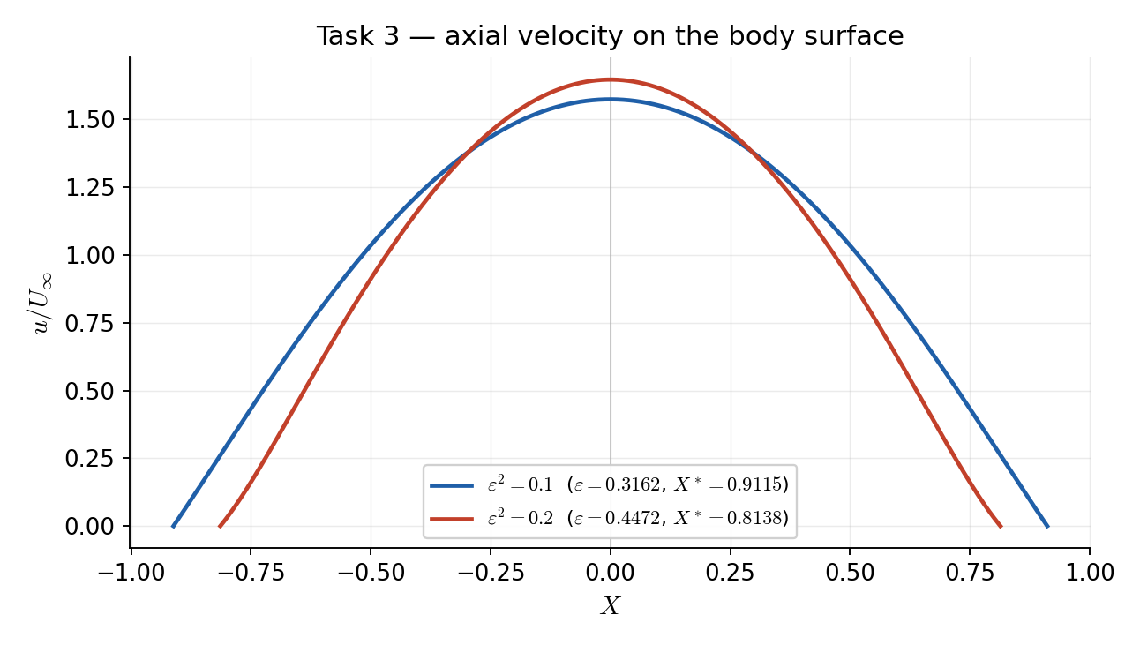}}
  \\
   \subfigure[Radial gas velocity component $v$ at surface..]{
  \includegraphics[height = 4.6cm]{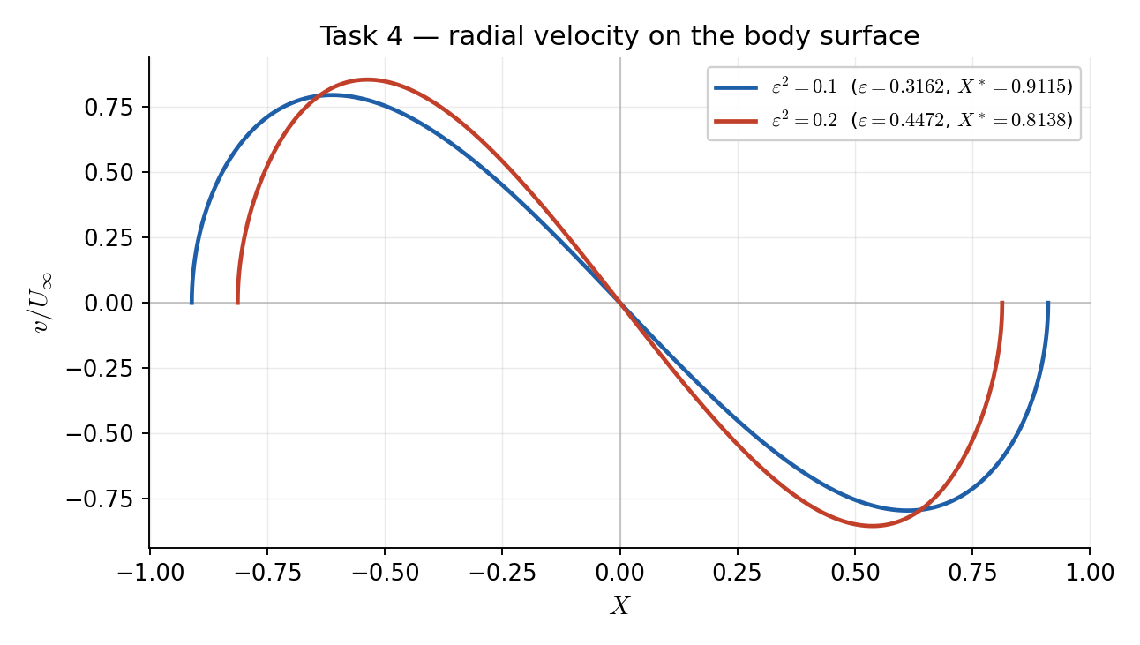}}
   \subfigure[Gas speed $U_s$ at surface.]{
  \includegraphics[height = 4.6cm]{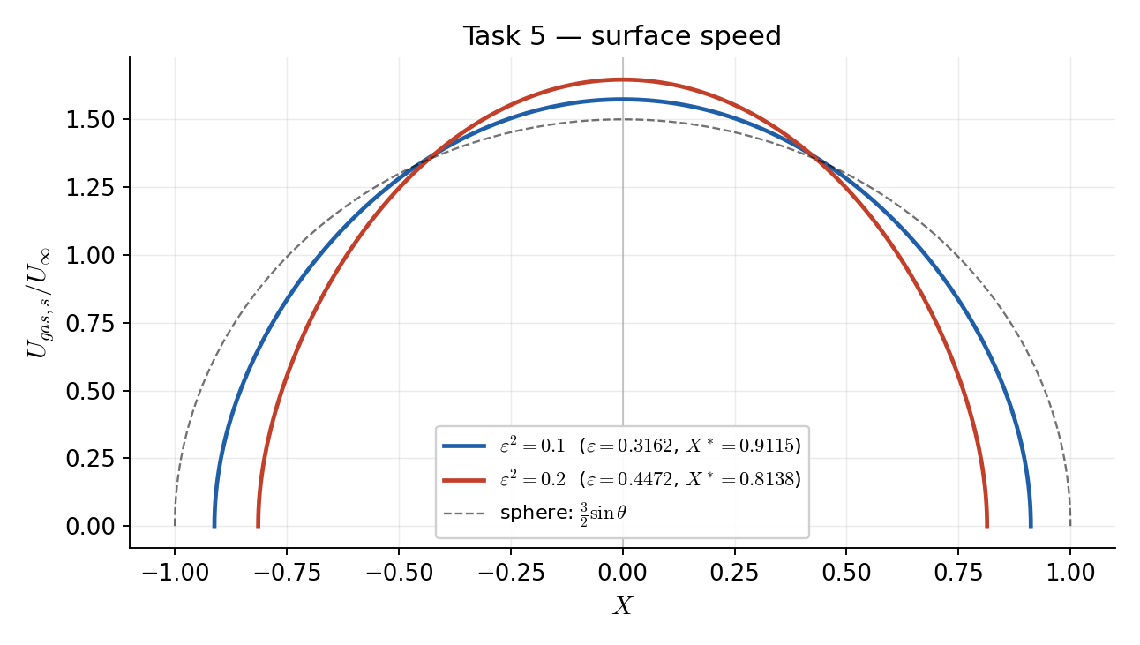}}
   \caption{For two selected values of $\varepsilon$, the normalized values for doublet-ring radius,  values at the droplet surface determined through Equation (\ref{Phi4}) are shown for velocity potential,  velocity components, and speed for the gas. Droplet shape is also shown. $\varepsilon^2$ values of $0.1$ and $0.2$ are used.}
  \label{potential}
  \end{figure}

\begin{figure}[thbp]
\subfigure[Streamlines for sphere with $\varepsilon^2 = 0.0$. ]{
  \includegraphics[height = 6.0cm]{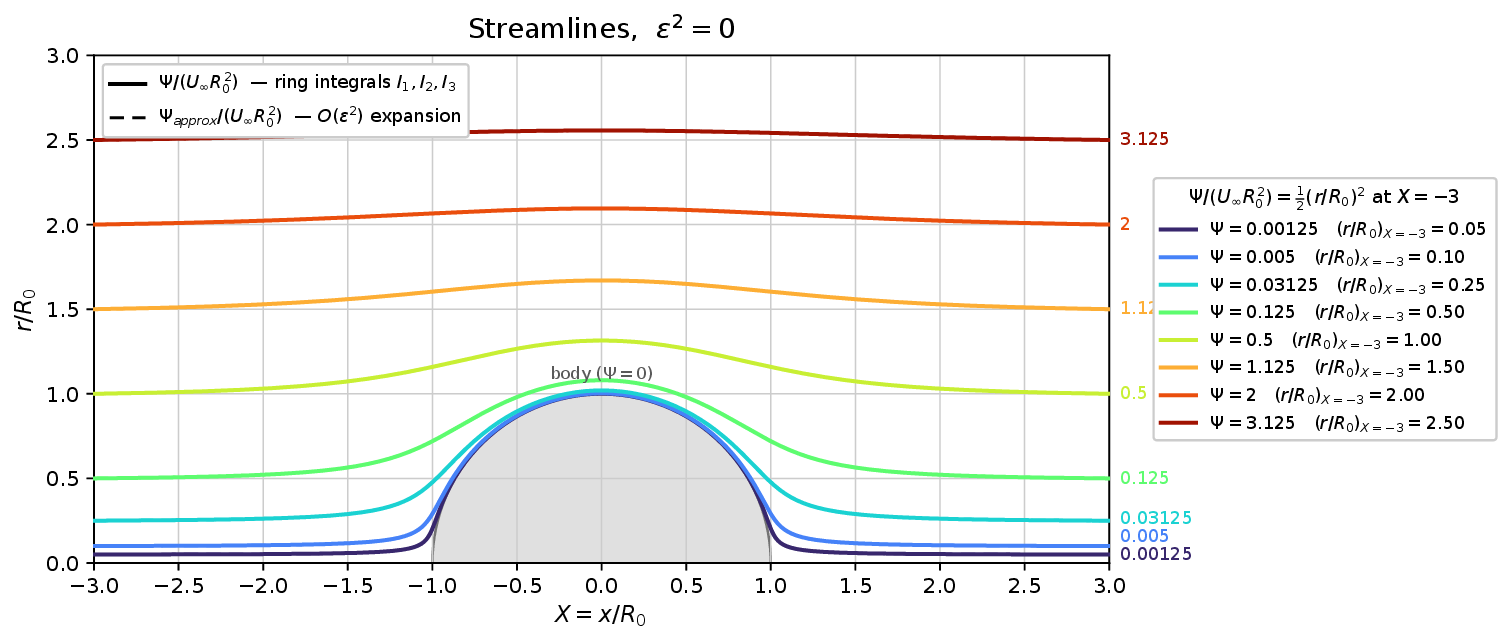}}
  \\
 \subfigure[Streamlines for $\varepsilon^2 = 0.1$. ]{
  \includegraphics[height = 6.0cm]{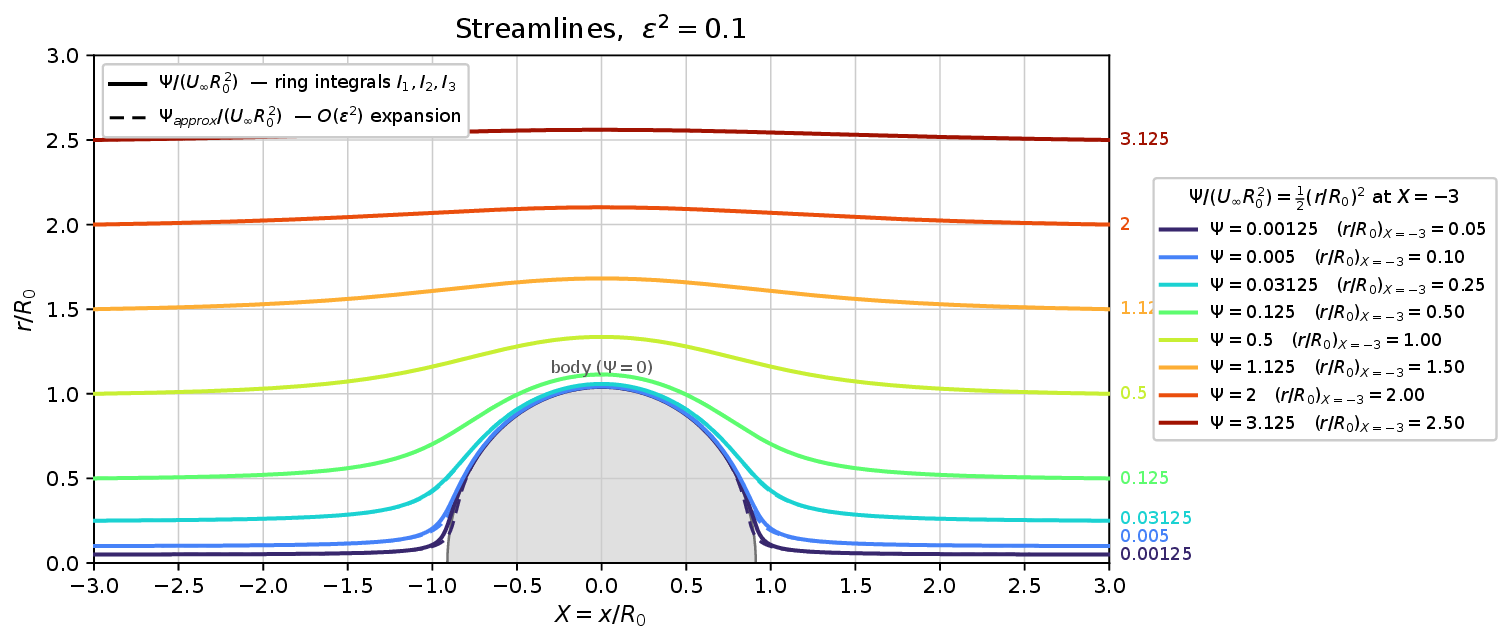}}
  \\
   \subfigure[Streamlines for $\varepsilon^2 = 0.2$.]{
  \includegraphics[height = 6.0cm]{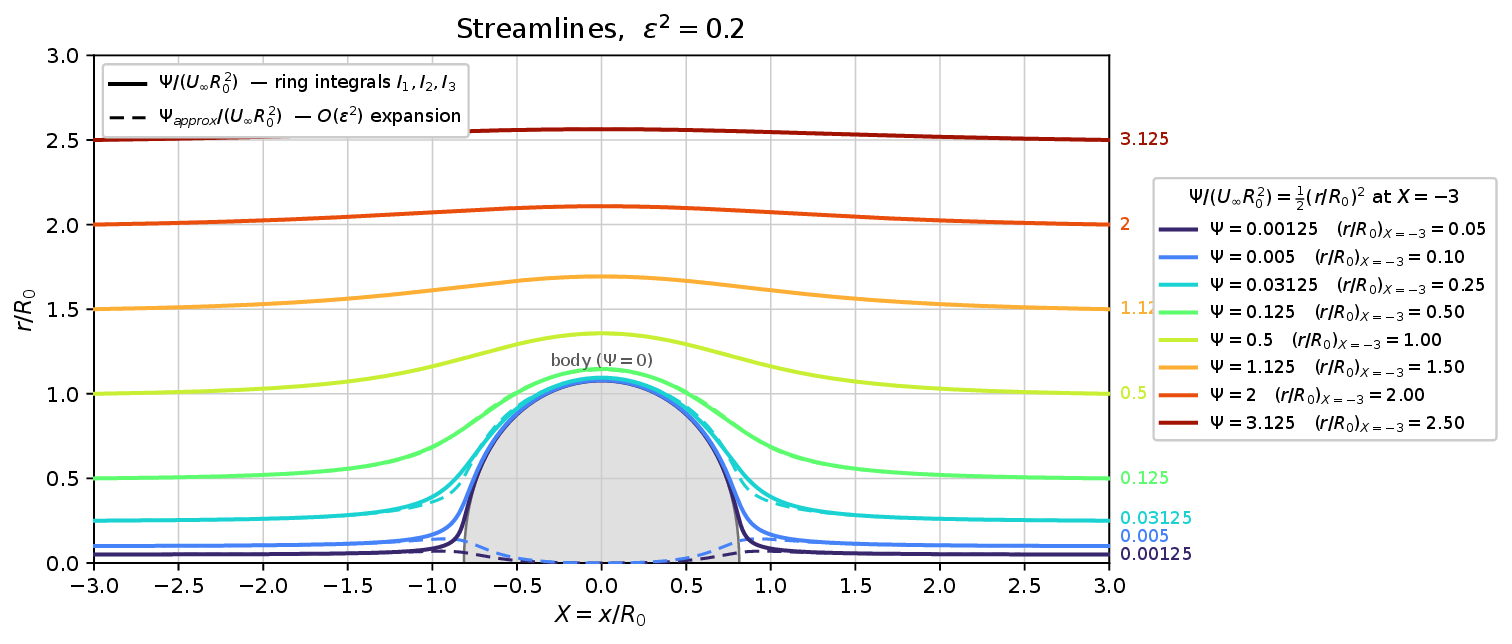}}
   \caption{For three selected values of $\varepsilon$, the normalized values for doublet-ring radius,  streamlines are shown for the flow around the oblate spheroid.  $\varepsilon^2$ values of $0.1$ and $0.2$ are used. Solid lines come from exact solution of potential flow using Equation (\ref{Phi4}) while dashed lines come from the perturbation theory using Equation (\ref{Phi5}). }
  \label{potential2}
  \end{figure}

Figure \ref{potential} shows some interesting results for the gas velocity along the droplet surface based on Equation (\ref{Phi4}) and the integration of the ordinary differential equation $dr/dx =v/u$ with the initial condition given as the stagnation point. Subfigure \ref{potential}b shows the deviation from the ellipsoidal shape given by the straight dashed lines. The $\varepsilon^2 =0$ choice gives exactly a sphere. The deviation from the ellipsoid increases with increasing doublet-ring normalized radius $\varepsilon$.
Figure \ref{potential2} shows as solid lines the streamlines determined by integration of $dr/dx = v/u$  given  initial points at $x=0$. The flow field remains qualitatively similar as $\varepsilon$ increases with symmetry maintained in the $x$-direction. The dashed lines will be explained later.

The  above integral forms would require substantial computational time if the droplet behavior were described in the sub-grid formulation for a  spray flow, where droplet velocity, size, and shape were changing with the Lagrangian time in the reference frame moving with the droplet. In search of a less costly, albeit more approximate, method, we now examine a perturbation approach. Specifically, we  expand the several integrands as Taylor series in powers of the small parameter $\varepsilon$, followed by the integration.
\begin{eqnarray}
I_1 &\equiv& \frac{1}{4 \pi}\int_0^{2\pi} \frac{d\alpha^*}{\Big[ \frac{r^2 +x^2}{R_0^2}\Big(1 + \frac{\varepsilon^2 R_0^2}{r^2 +x^2}- 2\varepsilon\frac{rR_0}{r^2 +x^2} \cos \alpha^* \Big) \Big]^{3/2}}
\nonumber \\
&=& \frac{1}{4 \pi} \Big(\frac{R_0^2}{r^2 + x^2}\Big)^{3/2}\int_0^{2\pi} \frac{d\alpha^*}{\Big[ 1 +\frac{3\varepsilon^2}{2} \frac{R_0^2}{r^2 +x^2}- 3\varepsilon\frac{rR_0}{r^2 +x^2} \cos \alpha^* +\frac{3\varepsilon^2}{2}\Big(\frac{rR_0}{r^2 +x^2} \Big)^2 \cos^2 \alpha \Big]}  +O(\varepsilon^3)
\nonumber \\
&=& \frac{1}{4 \pi} \Big(\frac{R_0^2}{r^2 + x^2}\Big)^{3/2}\int_0^{2\pi} \Big[ 1 -\frac{3\varepsilon^2}{2} \frac{R_0^2}{r^2 +x^2}+ 3\varepsilon\frac{rR_0}{r^2 +x^2} \cos \alpha^* +\frac{15\varepsilon^2}{2}\Big(\frac{rR_0}{r^2 +x^2} \Big)^2 \cos^2 \alpha \Big] d\alpha^*  +O(\varepsilon^3)
\nonumber  \\
&=& \frac{1}{2}\Big(\frac{ R_0^2}{r^2 + x^2} \Big)^{3/2} +\frac{\varepsilon^2}{8} \frac{R_0^5(9r^2 -6x^2)}{(r^2 +x^2)^{7/2}}    +O(\varepsilon^3)
\label{int1}
\end{eqnarray}
By similar perturbation expansions followed by integration over $\alpha^*$, we obtain
\begin{eqnarray}
I_2 &=& \frac{1}{2}\Big(\frac{ R_0^2}{r^2 + x^2} \Big)^{5/2} +\frac{\varepsilon^2}{8} \frac{R_0^7(25r^2 -10x^2)}{(r^2 +x^2)^{9/2}}    +O(\varepsilon^3)
\nonumber \\
I_3 &=& \frac{5\varepsilon}{4} \frac{ R_0^6 r}{(r^2 +x^2)^{7/2}}  + O(\varepsilon^3)
\label{int2}
\end{eqnarray}
Note that all three integrals and therefore the ring doublet go to zero value as $r^2$ and/ or $x^2$ become large, thereby yielding only the free stream flow.
Let us reconstruct the results from Equations (\ref{Phi4}, \ref{int1}, \ref{int2}) to show the perturbation effect more clearly. A point doublet of strength $U_{\infty}\tilde{d}\varepsilon^2$ is added at the origin for a reason to be fully discussed later. It will increase the semi-minor axis and reduce the semi-major axis while maintaining the liquid volume.  Here, we use the spherical radius $R= \sqrt{r^2 +x^2}$.
\begin{eqnarray}
\Phi &=& U_{\infty}\Big[x + (1 + \tilde{d}\varepsilon^2)\frac{x}{2}\Big(\frac{R_0}{R}\Big)^3  +\frac{\varepsilon^2 x}{8} \frac{R_0^5(9r^2 -6x^2)}{R^7} \Big] +O(\varepsilon^3)
\nonumber \\
u &=&  U_{\infty}\Big[ 1  +(1 + \tilde{d}\varepsilon^2)\frac{R_0^3(r^2 - 2x^2)}{2 R^5}   
 +
\varepsilon^2\frac{R_0^5\big(    \frac{9}{8}r^4 + 3 x^4-9 r^2 x^2\big)}{R^9} \Big]
+ O(\varepsilon^3)
\nonumber \\
v &=&  -\frac{3}{2} U_{\infty}\frac{xr R_0^3} {R^5}\Big[  1 +\tilde{d}\varepsilon^2
-\varepsilon^2\frac{R_0^2(10 x^2 - 15r^2/2)}{2R^4} \Big]
+O(\varepsilon^3)
\label{Phi5}
\end{eqnarray}
$u,I_1, I_2, $ and $I_3$ are  symmetric in $x$, while $\Phi$ and $v$ are anti-symmetric. Stagnation points are predicted with $\tilde{d} = 1/2$ at $r =0, x/R_0 = 1- 5\varepsilon^2/6$ and $r=0, x/R_0 = -1 +5 \varepsilon^2/6$. 

\begin{figure}[thbp]
 \subfigure[Constant potential lines for $\varepsilon^2 = 0$, i.e., a sphere. ]{
  \includegraphics[height = 6.0cm]{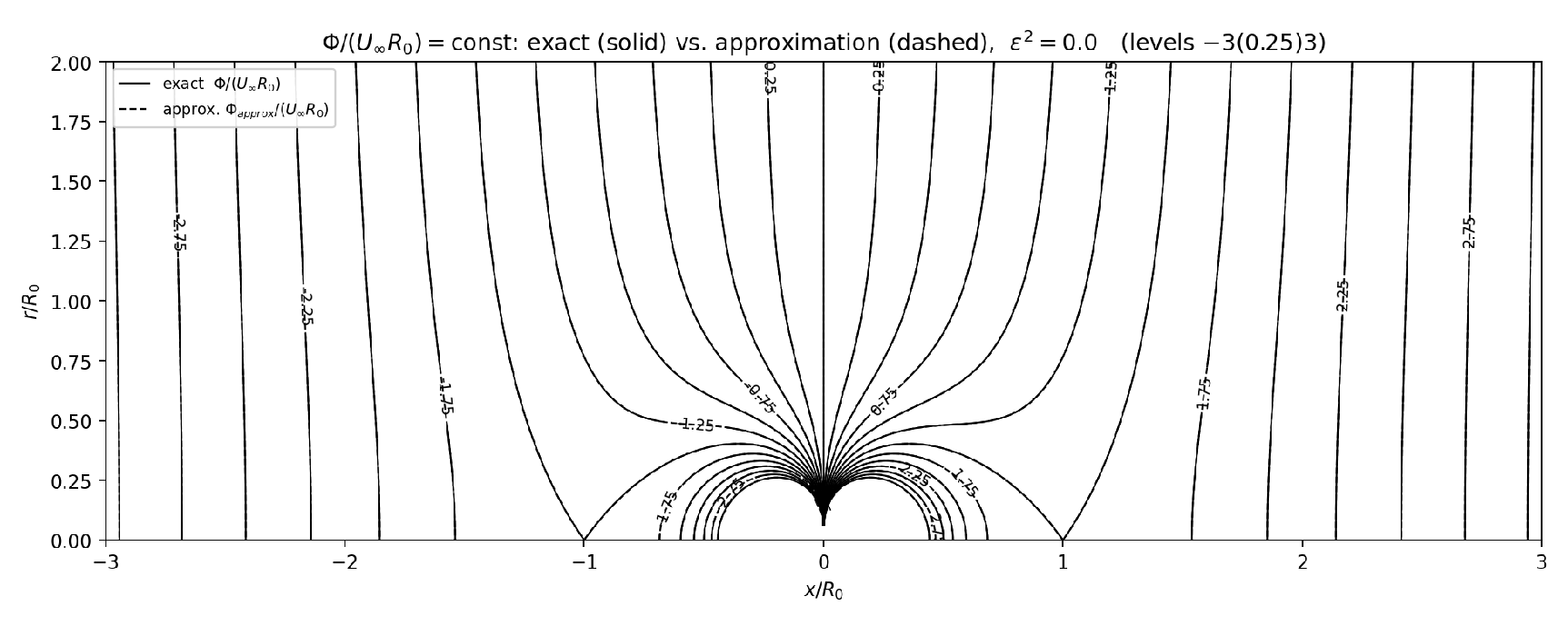}}
  \\
   \subfigure[Constant potential lines for $\varepsilon^2 = 0.1$.]{
  \includegraphics[height = 6.0cm]{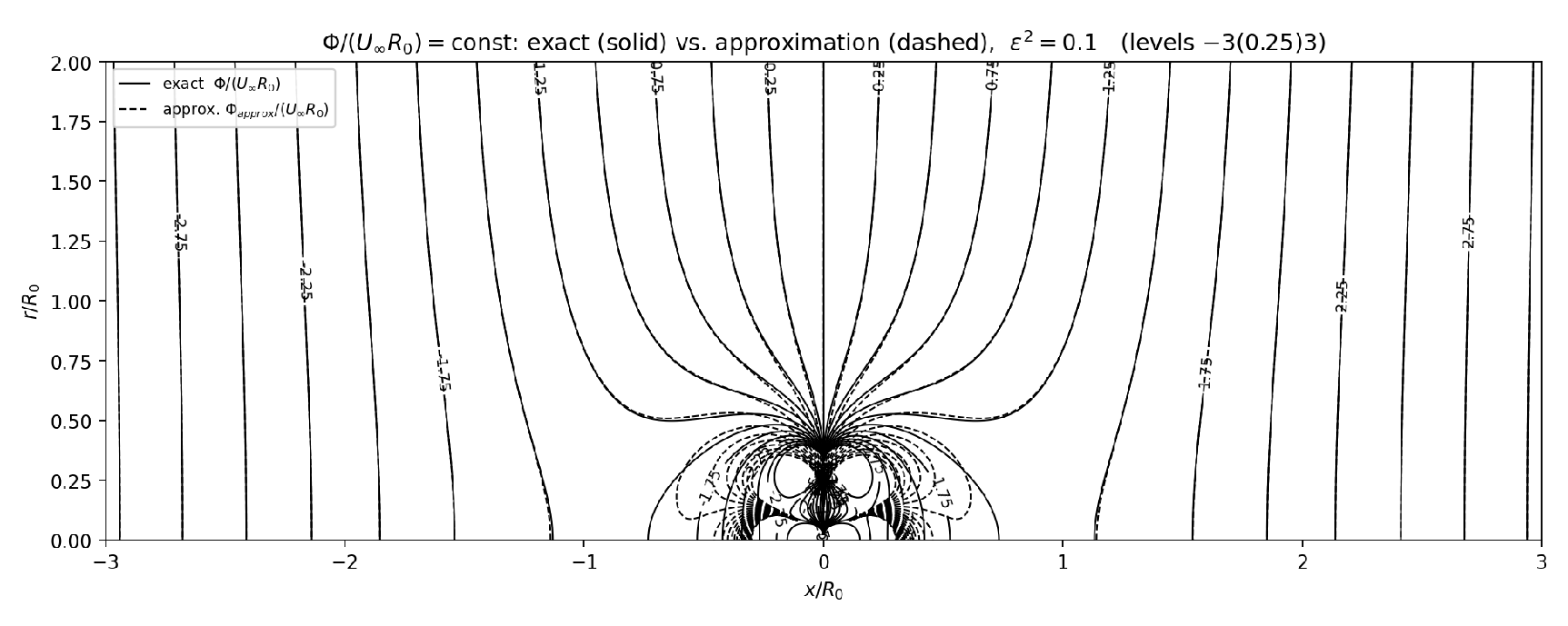}}
  \\
  \subfigure[Constant potential lines for $\varepsilon^2 = 0.2$.]{
  \includegraphics[height = 6.0cm]{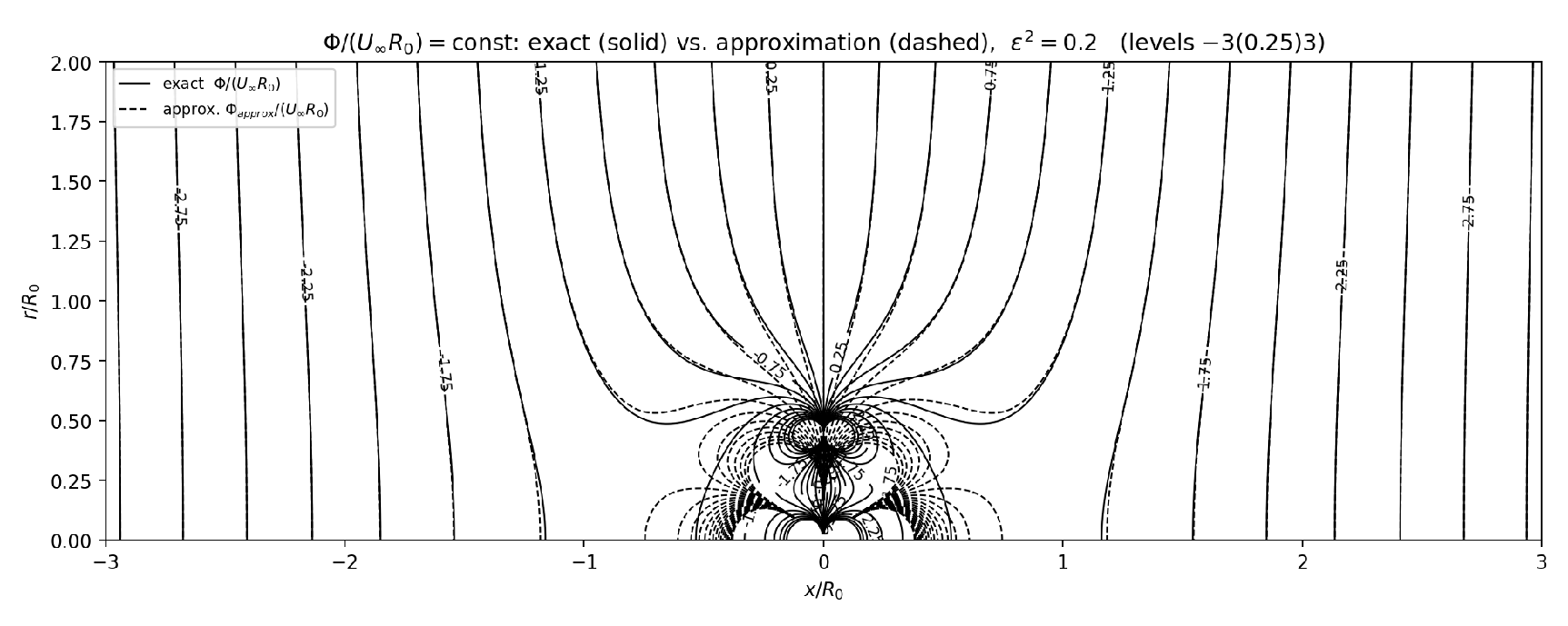}}
   \caption{For three selected values of $\varepsilon$, the normalized values for doublet-ring radius,  solid lines of constant potential values predicted by exact theory using Equation (\ref{Phi4}) are compared with dashed lines of constant potential values predicted by perturbation theory using Equation (\ref{Phi5}) for the flow around the oblate spheroid.  $\varepsilon^2$ values of $0, 0.1$ and $0.2$ are used.}
  \label{potentialcompare}
  \end{figure}

Comparisons of the velocity potential results for the ring doublet alone from the exact analysis and the perturbation analysis are provided in Figure \ref{potentialcompare}. The results show both the flow in the gas flow domain and in the image region inside the droplet. For the sphere with $\varepsilon =0$, the perturbation theory exactly agrees exactly with the original theory. The agreement for small values of $\varepsilon^2$  is quite excellent in the region of physical interest, approximately $x^2 +r^2 \geq R_0^2$. In the spherical case, only a point doublet at the origin remains while the effect of the ring doublet is shown in subfigures \ref{potentialcompare}b and \ref{potentialcompare}c for the spheroidal cases. Actually, as $\varepsilon \rightarrow 0$, the ring doublet collapses to strong point doublet at the origin and the previously added weak point doublet disappears. The difference between the behaviors for the spherical and spheroidal cases depends on the square of the radius of the ring doublet, which in many interesting cases may be considered as  a perturbation on the spherical case.

Using the above results and setting  $u= (1/r) \partial \Psi/\partial r$ and $v = -(1/r) \partial \Psi/ \partial x$, a stream function $\Psi$ can be presented after some integration.
\begin{eqnarray}
    \frac{\Psi}{U_\infty R_0^2} = \frac{r^2}{2R_0^2} -\frac{r^2R_0}{2(x^2 +r^2)^{3/2}}\Big(1 + \tilde{d}\varepsilon^2 \Big)  + \frac{3\varepsilon^2 r^2 R_0^3}{2}\Big[ \frac{1}{(x^2 + r^2)^{5/2}} - \frac{5r^2}{4(x^2 + r^2)^{7/2}}   \Big] +O(\varepsilon^3)
    \label{perturbstream}
\end{eqnarray}
Interestingly, we may develop the same form as the ring doublet portion by using perturbation theory on the vortex ring formula
\citep{Saffman, Fraenkel1972, Norbury1972}. Consider an infinitesimally   thin circular vortex ring with delta function strength and ring radius $\sigma$ in the $x=0$ plane. Use that radius as  a perturbation parameter (i.e., $\varepsilon$ in nondimensional form) and expand to give a series in powers of the parameter. The first two terms in the expansion match the ring doublet contribution shown in Equation (\ref{perturbstream}).

A streamline can be determined by $dr/dx = v/u$; so, we can expect symmetry in $x$ for the stream function. 
Figure \ref{potential2} shows as dashed lines the results from using the velocity given by Equation (\ref{Phi5}) based upon perturbation theory. The curves deviate more the exact theory as $\varepsilon$ grow and as the stagnation points are approached. Generally, the agreement is good and an exact match occurs as $\varepsilon \rightarrow 0$.
\begin{figure}[thbp]
 \subfigure[Constant stream function for $\varepsilon^2 = 0.1 , \tilde{d} =0$. ]{
  \includegraphics[height = 4.6cm]{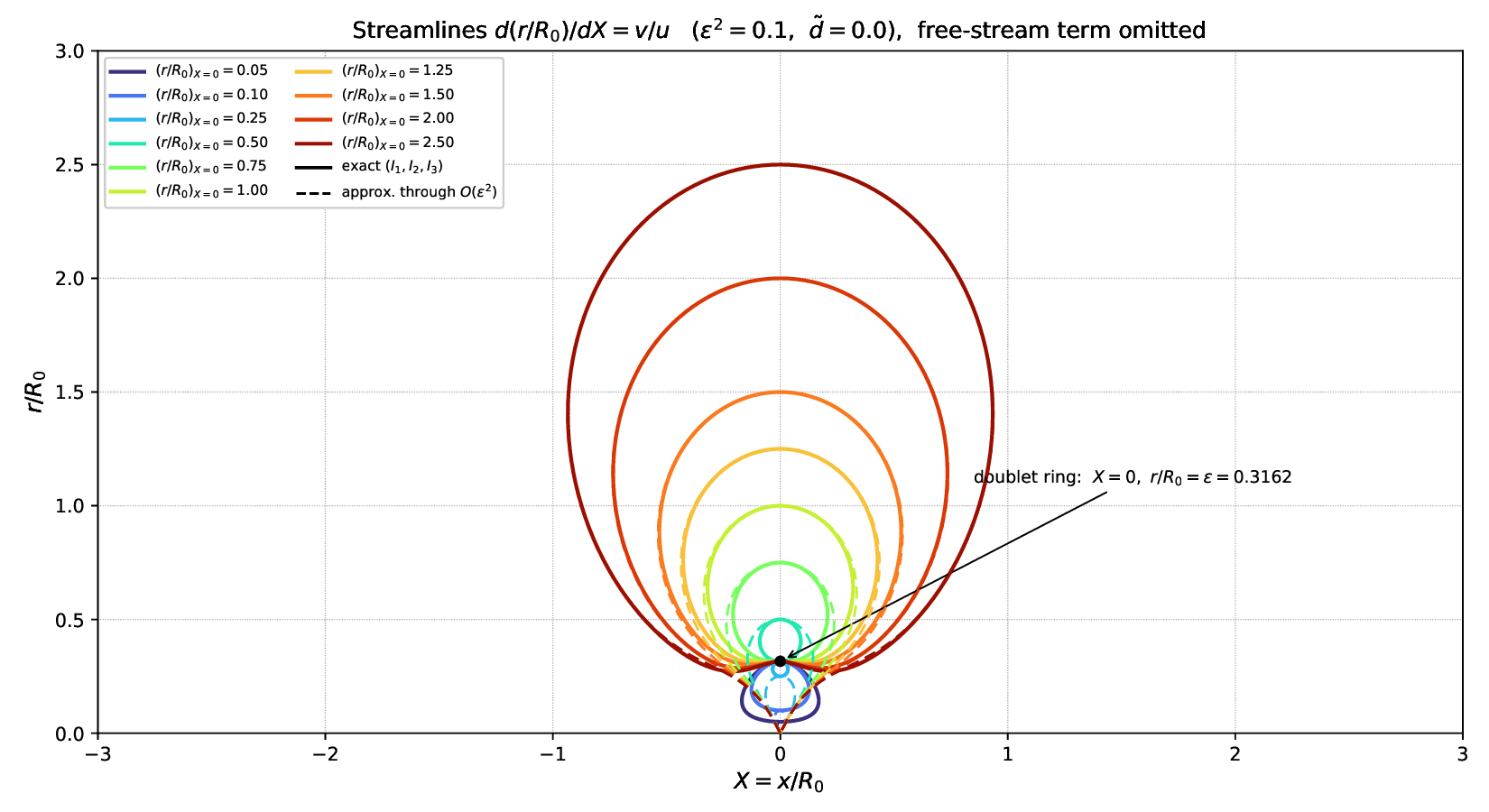}}
  \\
   \subfigure[Constant stream function for $\varepsilon^2 = 0.1 , \tilde{d} =0.5$.]{
  \includegraphics[height = 4.6cm]{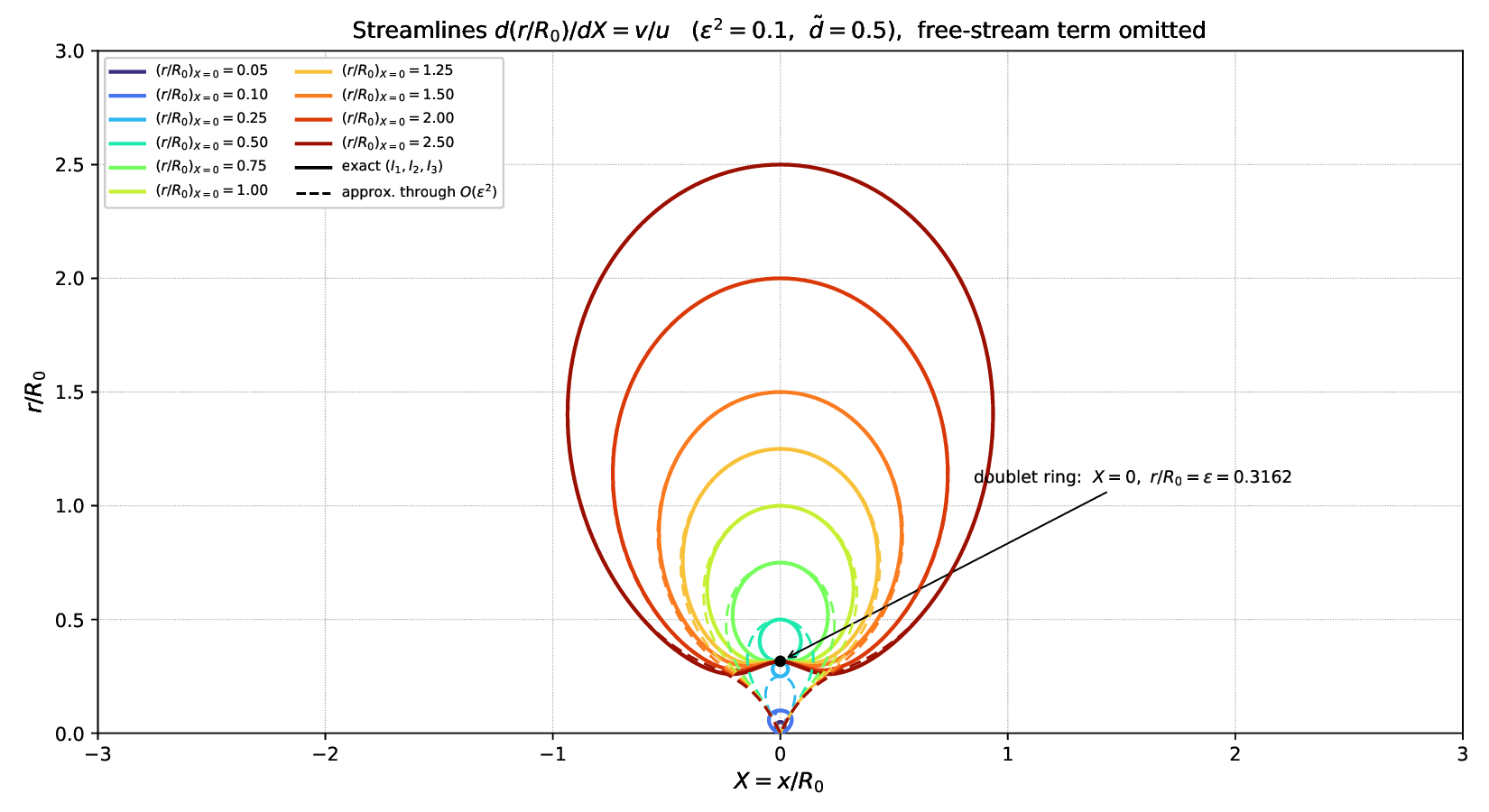}}
  \\
  \subfigure[Constant stream function for $\varepsilon^2 = 0.2 , \tilde{d} =0$.]{
  \includegraphics[height = 4.6cm]{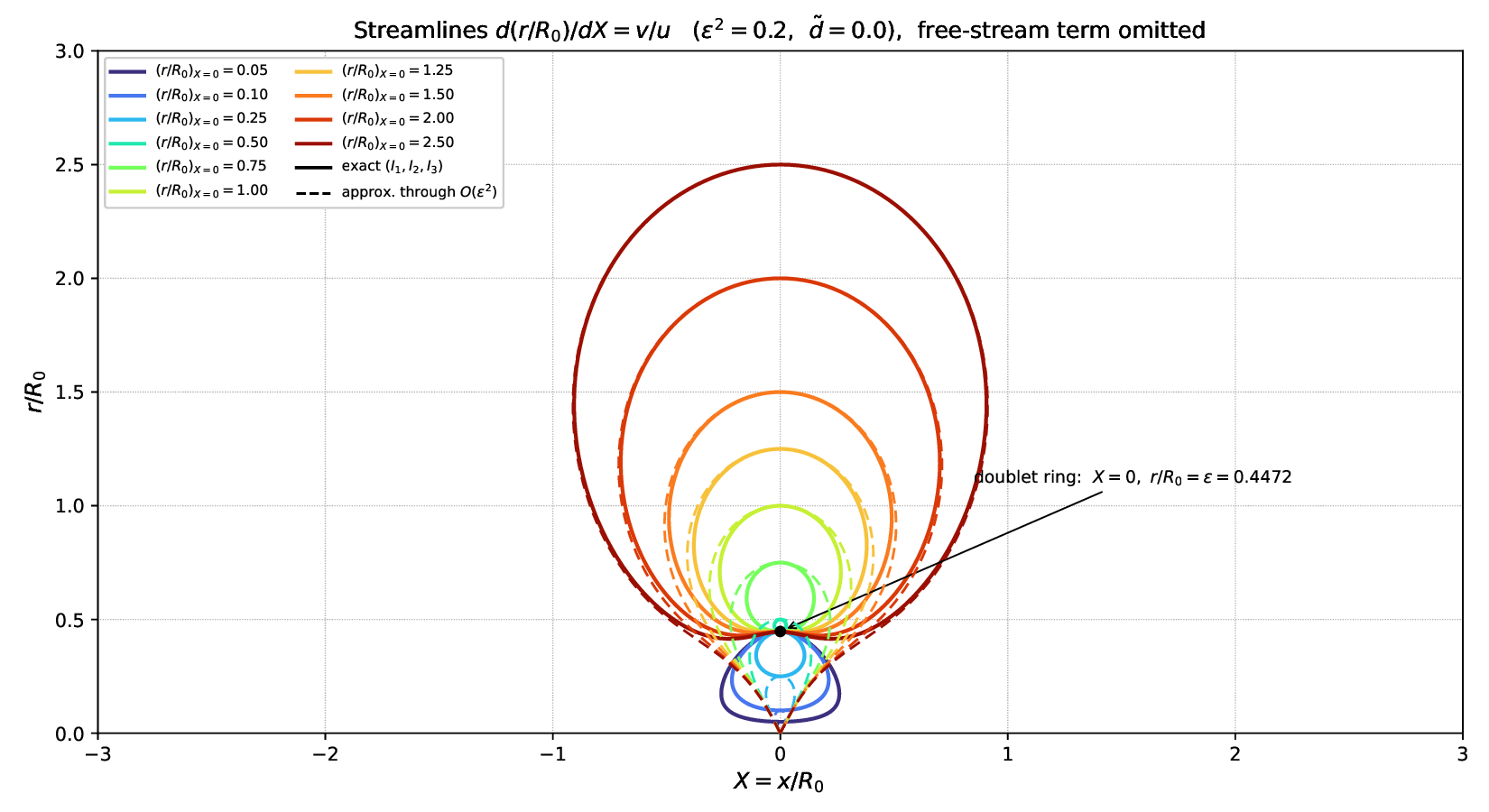}}
  \\
  \subfigure[Constant stream function for $\varepsilon^2 = 0.2 , \tilde{d} =0.5$]{
  \includegraphics[height = 4.6cm]{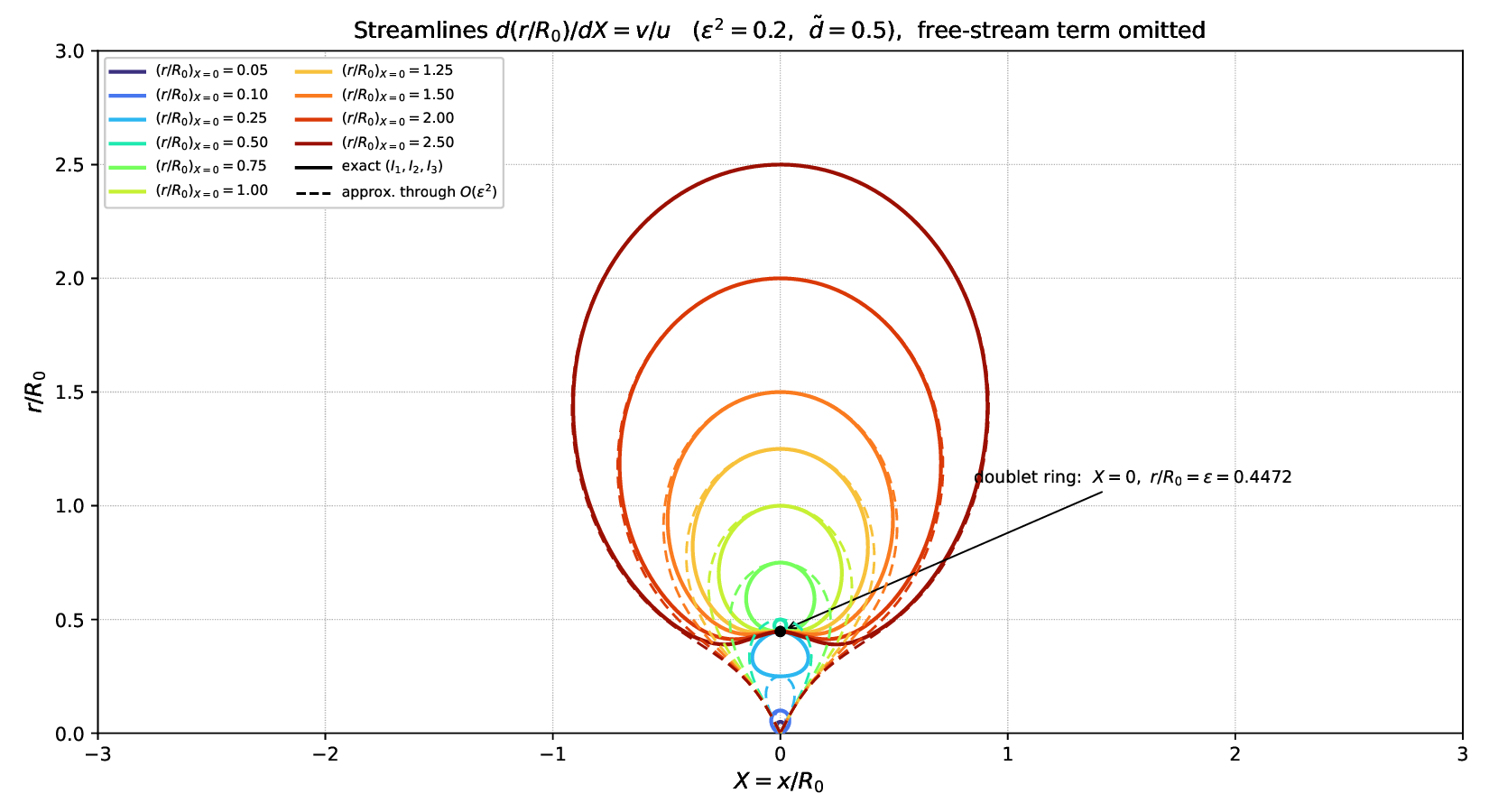}}
   \caption{For two  values of $\varepsilon$ and two values of $\tilde{d}$ with omission of the free stream and focus on the doublets,  solid lines of constant stream function values predicted by exact theory from Equation (\ref{Phi4}) are compared with dashed lines of constant stream function values predicted by perturbation theory using Equation (\ref{Phi5}). }
  \label{doubletcompare}
\end{figure}

Let us examine the streamline shapes, especially for the droplet surface. 
Variables are converted to non-dimensional forms and we use the squares of the two independent variables: $\xi=x^2/R_0^2, \eta = r^2/ R_0^2, R^2/R_0^2 = \tilde{R}^2 = \xi + \eta$.
We will start with $rdr/(xdx) = d\eta/d\xi = rv/xu$, substitute from Equation (\ref{Phi5}), and  properly order the series in $\varepsilon$.
\begin{eqnarray}
\frac{d\eta}{d\xi}= - \frac{3\eta}{2\tilde{R}^5}\Bigg[  \frac{   1 +
\tilde{d}\varepsilon^2
-\varepsilon^2\frac{(10 \xi - 15\eta/2)}{2\tilde{R}^4}}{1  +\frac{(\eta - 2\xi)}{2 \tilde{R}^5}   
+\varepsilon^2\frac{\tilde{d}}{2\tilde{R}^5}(\eta-2\xi) +
\varepsilon^2\frac{\frac{9}{8}\eta^2 + 3 \xi^2-9 \eta\xi}{\tilde{R}^9}} \Bigg]  +O(\varepsilon^3)
\label{gasstream}
\end{eqnarray}
Actually, we expect the error to be $O(\varepsilon^4)$ but we do not need to prove and use it now. Equation (\ref{gasstream}) can determine $\eta(\xi)$   for any streamline. For example, at $\xi =0$, a value of $\eta$ can be chosen as an initial condition for the ordinary differential equation that also distinguishes the particular streamline.  We will now focus only on the streamline that describes the droplet boundary. Consider $\eta =\eta_0 +\varepsilon^2 \eta_1 + O(\varepsilon^3)$. Substitution and separation by powers of $\varepsilon$ yields to lowest order
\begin{eqnarray}
\Big[ (\eta_0 +\xi)^{5/2} +\frac{\eta_0 - 2\xi} {2} \Big]\frac{d\eta_0}{d\xi} &=& -\frac{3\eta_0}{2}  \label{streamzero}
\end{eqnarray}

The solution $\eta_0 = 1 - \xi$, describing a circle  is readily seen. Implicitly, the boundary condition $\eta_0(0) = 1$ is used which brings the solution to apply to the surface streamline. Using $\tilde{R}_0^2 = 1$, we may simplify the equation governing $\eta_1$. Now,
\begin{eqnarray}
\Big[ (\eta_0 +\xi)^{5/2} +\frac{\eta_0 - 2\xi} {2} \Big]\frac{d\eta_1}{d\xi}
+\big[   \frac{\tilde{d}}{2}(\eta_0-2\xi) +
\frac{9}{8}\eta_0^2 + 3 \xi^2-9 \eta_0\xi \big] \frac{d\eta_0}{d \xi}&=& -\frac{3\eta_1}{2}   -    \frac{3 \eta_0\tilde{d}}{2}
+ \frac{(15 \xi\eta_0 - 45\eta_0^2/4)}{2} \nonumber \\ 
&-& \frac{5(\eta_0 + \xi)^{3/2}}{2}\frac{d\eta_0}{d\xi}\eta_1
- \frac{\eta_1}{2}\frac{d\eta_0}{d\xi} 
\label{eta1}
\end{eqnarray}
Substitution of the $\eta_0$ solution into the $\eta_1$ equation yields for the droplet boundary perturbation ordinary differential equation and its solution 
\begin{eqnarray}
(1 - \xi) \frac{d \eta_1}{d \xi} &=&   \eta_1 -\frac{2\tilde{d}}{3}  + 2 - 5(1 -\xi)  \;\; ; \;\;\nonumber \\
\eta^*(\xi) &\equiv & \eta_1(\xi) - \frac{2\tilde{d}}{3} +2  \;\; ;\;\;
\frac{d (\eta^*(1-\xi))}{d\xi} = -5(1-\xi)  \;\; ; \;\;\nonumber \\
\eta^* &=& \frac{C}{\xi -1} +\frac{5(1 - \xi)}{2}  \;\; ; \;\; \eta_1 = \frac{2\tilde{d}}{3}  - 2 + \frac{C}{\xi -1} +\frac{5(1 - \xi)}{2} 
\end{eqnarray}
We choose $\tilde{d} = \frac{3\eta_1(0)}{2} -\frac{3}{4}  $; consequently, $C=0$ and the singular behavior is removed to yield 
\begin{eqnarray}
    \eta_1(\xi) &=& \eta_1(0) - \frac{5\xi}{2}  \;\; ; \;\; \nonumber \\
    \eta(\xi) &=& \eta_0(\xi) + \varepsilon^2\eta_1(\xi) + O(\varepsilon^3) = 1 +\varepsilon^2 \eta_1(0) - (1 + \frac{5\varepsilon^2 }{2} )\xi + O(\varepsilon^3) \;\; ; \;\; \nonumber \\ 
    1 &=&  \frac{\eta}{1+\varepsilon^2 \eta_1 (0)}  +  \frac{\xi}{1 +\varepsilon^2(\eta_1 (0) -\frac{5}{2} )}   + O(\varepsilon^3)
    \label{gasellipse}
\end{eqnarray}
 Neglecting the higher order terms, we have an ellipsoid with semi-major axis $ a = \sqrt{1+\varepsilon^2 \eta_1 (0)}$ and semi-minor axis $ b = \sqrt{1 +\varepsilon^2(\eta_1 (0) -\frac{5}{2} )}$. The addition of the point doublet at the origin has not only removed the singularity but has also forced the ellipsoidal shape. The value of $\eta_1(0)$ can be chosen to match the particular ellipsoidal shape found in other portions of the analysis.  In retrospect after the perturbation analysis, we did return to the formulation of Equation (\ref{Phi4})
 and added a point doublet of magnitude  $(\tilde{d}\varepsilon^2 U_{\infty}R_0^3/2) x/[(r/R_0)^2 +X^2]^{3/2}$ to the flow to adjust towards the ellipsoidal droplet shape. It  results in an increase in the semi-major axis with the corresponding decrease in the semi-minor axis.

The perturbation theory does result in errors of $O(\varepsilon^4)$ which in a relative sense become significant near the stagnation points.  Figure \ref{potential2} shows streamlines calculated from both the exact potential theory and the perturbation theory. The differences in our $\varepsilon$ range are minor away from the droplet surface. They are tolerable in the high-speed region near $x=0$, but become more severe near the stagnation points because there are differences in the predicted locations of the stagnation points.
In Figure \ref{potentialcompare}, the potential lines are shown for both the exact theory and the perturbation theory. The match is excellent at distance from the droplets but is weaker near the droplet. We show the potential lines inside the droplet, although they are unphysical, in order to explain the mathematical behavior. As $\varepsilon$ increases, the ring-doublet radius increases which moves the shape further from a sphere. The weak doublet at the origin gets stronger, assuring the $x$-width of the body remains largest at $r=0$.

 Figure 7 shows  the impacts of the portion of the potential function for the ring doublet imposed in the $x=0$ plane and the weak point doublet at the origin. The free stream effect is eliminated in the plot. Comparisons are made at two values of the normalized ring doublet radius $\varepsilon = \sigma/R_0 = \sqrt{0.1}, \sqrt{0.2}$ with and without the weak doublet at the origin. Streamlines from the exact theory (solid lines) are compared to streamlines from perturbation theory (dashed lines) by taking pairs that cross at the same $x$ value at a chosen radial $r$ value, slightly greater than the ring radius.  When $\tilde{d} = 0$, the point doublet is eliminated.  $\tilde{d} =0.5$ is the value used in our prior calculations; the justification will be given in the next section.  The point doublet has a significant effect near the origin and near $r=0$ despite its low strength. As the  ring doublet radius increases, the streamlines tend to slope upwards a greater rate, expainng how the body would become more oblate. The case of $\varepsilon =0$ is not shown; the weak doublet disappears, the ring doublet collapses to a point doublet at the origin, and the difference between exact theory and perturbation theory disappears.

\begin{figure}[thbp]
 \subfigure [Droplet shape $r(X)$ . ]{
  \includegraphics[height = 4.6cm]{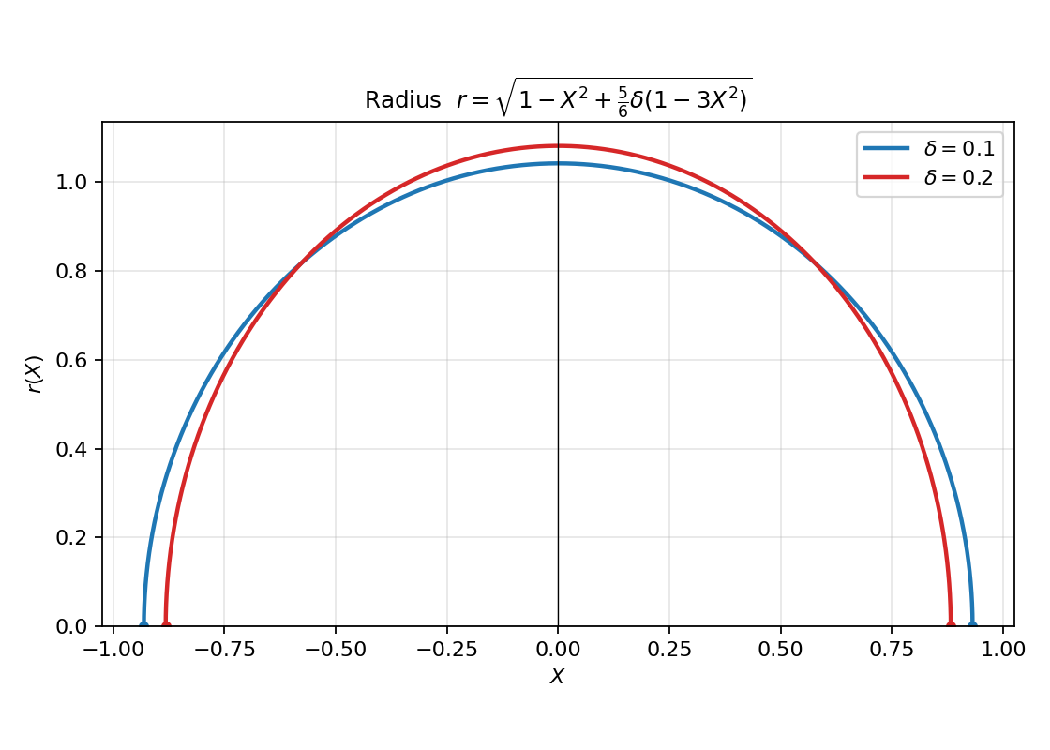}}
  \subfigure [Gas speed at droplet surface $U_s/U_{\infty}$. ]{
  \includegraphics[height = 4.6cm]{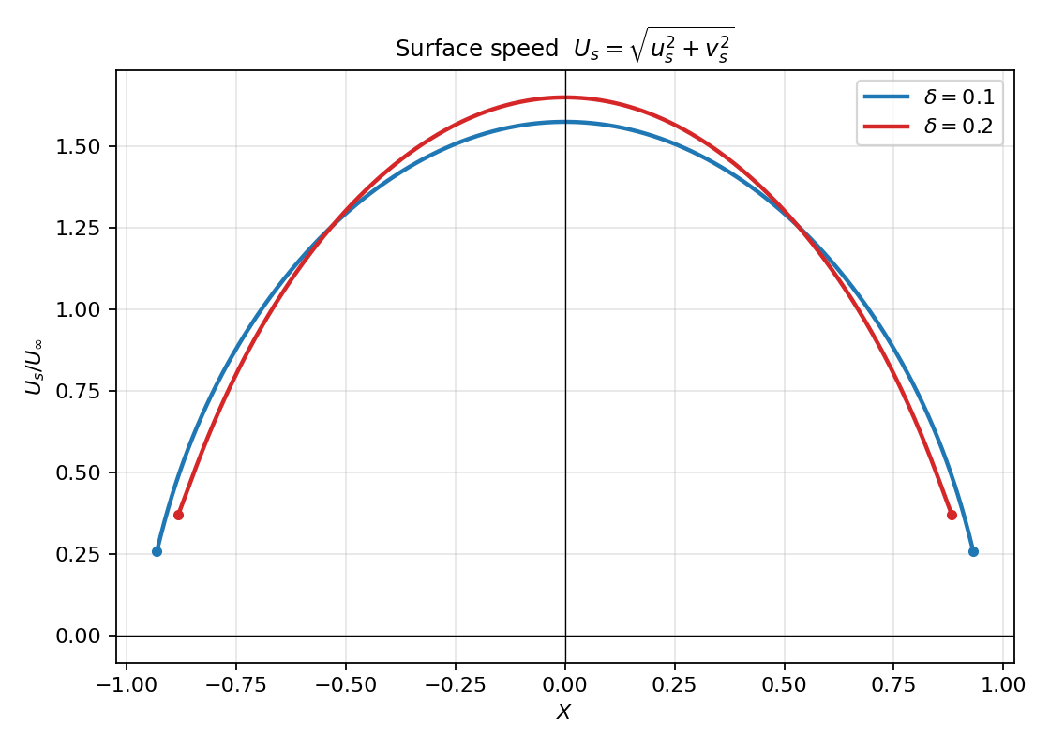}}
\\
   \subfigure[Radial gas velocity component $v/U_{\infty}$ at surface.]{
  \includegraphics[height = 4.6cm]{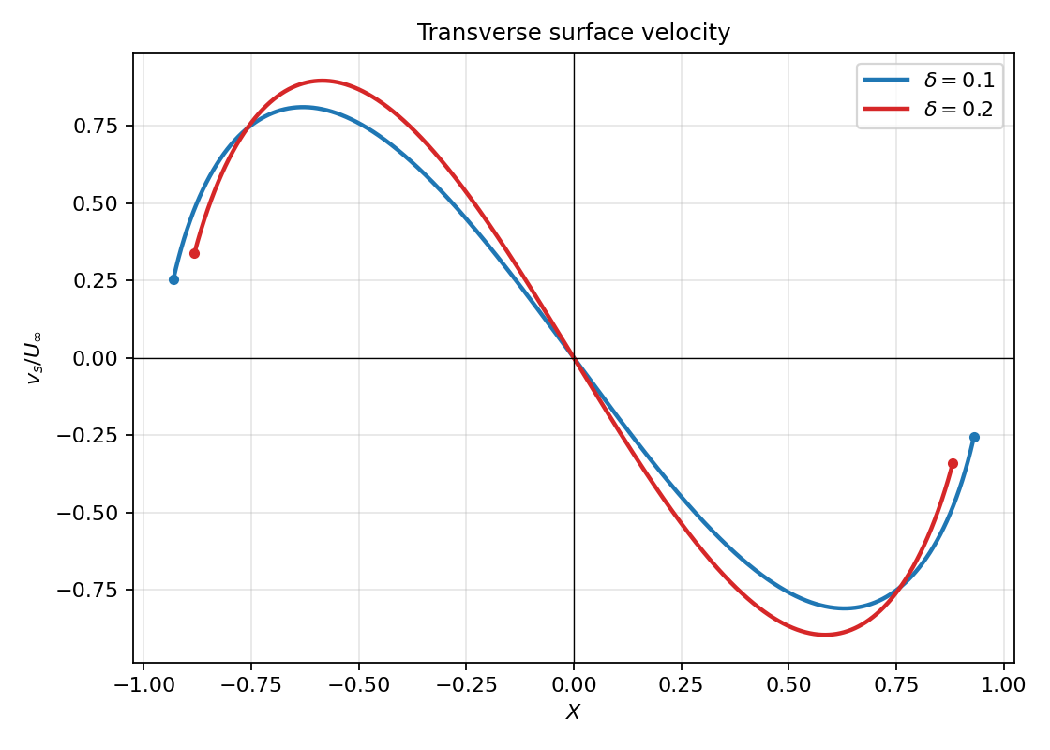}}
   \subfigure[Axial gas velocity component $u/U_{\infty}$ at surface.]{
  \includegraphics[height = 4.6cm]{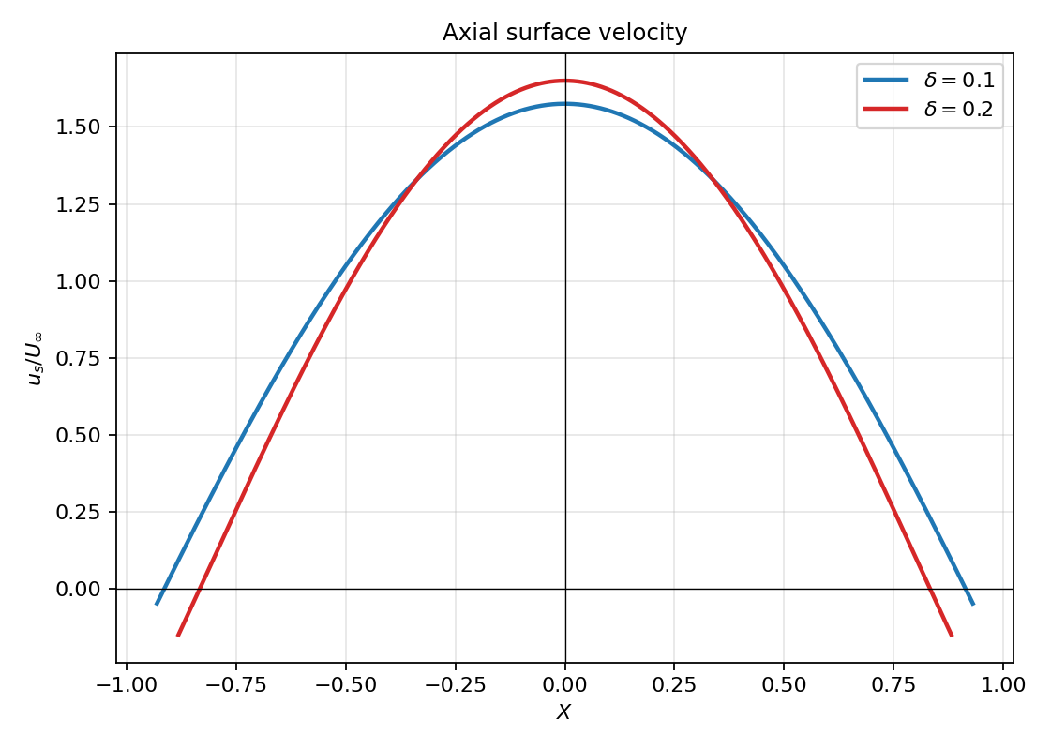}}
   \caption{For two selected values of $\varepsilon$ , the normalized values for doublet-ring radius, perturbation-theory results from Equations (\ref{surface}, \ref{speed2}) at the droplet surface are shown for velocity potential,  velocity components, and speed for the gas. Droplet shape is also shown. $\varepsilon$ values of $0.1$ and $0.2$ are used.}
  \label{perturbationfig}
  \end{figure}

Here, in Figure \ref{perturbationfig} we show the surface velocity and droplet shape determined by  the perturbation theory through Equations (\ref{Phi5}) and (\ref{gasellipse}).

\section{Connection of the External Potential Flow, Surface Tension, and Internal Circulation} \label{integrate}

We can match the semi-major and semi-minor axis from the three analyses.
The liquid-flow analysis and gas-flow analysis each gave an ellipsoid to the order of magnitude addressed, while we have an approximate ellipsoidal from the surface curvature analysis. Set $a = 1 + \varepsilon^2 \eta_1(0)/2 = 1 + 5\delta/12$ for the semi-major axis.
Thus, $\varepsilon = \sqrt {\delta} = \sqrt{We}/2; \eta_1(0) = 5/6; a = 1 + 5\delta/12 = 1 + 5We/48$. For the semi-minor axis, set $b = 1 + \varepsilon^2 (\eta_1(0) -  5/2)/2 \approx 1 - \delta \beta(\delta)/2  $. This yields  $b =1 - 5 We/24$. The fit with the $\beta$ variable cannot be perfect. Our ellipsoid is now described, after neglect of $O(\delta^2)$, as 
\begin{eqnarray}
1 = \frac{\eta}{1 + \frac{5\delta}{6}}  +\frac{\xi}{1 - \frac{5 \delta}{3}} +O(\delta^2)
= \frac{G^2}{(1 + \frac{5 \delta}{12})^2}  +\frac{X^2}{(1 - \frac{5 \delta}{6})^2} +O(\delta^2)
\approx \frac{G^2}{(1 + \frac{5 We}{48})^2}  +\frac{X^2}{(1 - \frac{5 We}{24})^2}
\label{surface}
\end{eqnarray}
Based on the value of $\eta_1(0)$, we can set the strength of the corrective doublet strength to be $\tilde{d}= 1/2$. Actually, this finding preceded and guided our earlier choice for  the point-dipole strength in the exact formulation of Equation (\ref{Phi4}).  It also was used in the calculations from perturbation theory for Figures \ref{potential2}, \ref{potentialcompare}, \ref{doubletcompare}, and \ref{perturbationfig}.

The surface velocity can be determined using Equations (\ref{Phi5}) and (\ref{surface}).
\begin{eqnarray}
 \eta &=& 1 -\xi + \delta \frac{5}{6}(1 - 3\xi) +O(\delta^2) = r^2 =  1 -X^2 + \delta \frac{5}{6}(1-3X^2) + O(\delta^2)
 \nonumber \\
\frac{u_s}{U_{\infty}}&=& 1 + \frac{1 -3X^2 + \delta \frac{5}{6}(1-3X^2)}{2\big(1 + \delta \frac{5}{6}(1-3X^2)\big)^{5/2}}  +\frac{\delta}{4}(1 - 3X^2) + \delta\big( \frac{9}{8}(1 -X^2)^2 +3X^4 -9(X^2 - X^4)  \big)  +O(\delta^2)  \nonumber \\
&=& 1 + \frac{1 -3X^2}{2} +\delta\Big( \frac{2(1 -3X^2)}{3} - \frac{25(1 -3X^2)^2}{24} +\frac{9}{8} - \frac{45}{4}X^2  + \frac{105}{8}X^4\Big) + O(\delta^2)  \nonumber \\
&=& 1 + \frac{1 -3X^2}{2} 
+\delta\Big( \frac{3}{4} -7X^2 + \frac{15}{4}X^4  \Big)   + O(\delta^2)  \nonumber \\
\frac{v_s}{U_{\infty}} &=&  -\frac{3}{2} \frac{X (1 -X^2 + \delta \frac{5}{6}(1 - 3X^2))^{1/2} } {\big(1 + \delta \frac{5}{6}(1-3X^2)\big)^{5/2}}\Big[  1 +\frac{\delta}{2}
-\delta\frac{(35 X^2 - 15)}{4} \Big]  +O(\delta^2)    \nonumber\\
 &=&  -\frac{3}{2} X (1 -X^2)^{1/2} \Big[  1 +\frac{\delta}{2}  +  \delta \frac{5}{12}\frac{(1 - 3X^2)}{1-X^2} -  \delta \frac{25}{12}(1-3X^2)
-\delta\frac{(35 X^2 - 15)}{4} \Big]  +O(\delta^2)     \nonumber\\
 &=&  -\frac{3}{2} X (1 -X^2)^{1/2} \Big[  1 +\frac{13\delta}{6}  +  \delta \frac{5}{12}\frac{(1 - 3X^2)}{1-X^2} -  \delta \frac{5}{2}X^2\Big]  +O(\delta^2) 
 \label{speed2}
\end{eqnarray} 
The maximum surface velocity on the gas side of the droplet surface is $U_{g, max} = (3U_{\infty}/2)(1 +\delta/2) +O(\delta^2)$ and occurs at $X=0$. From Equation (\ref{maxvelocity}), we infer that $U_{l,max} = \sqrt{\rho_g/\rho_l}U_{g,max} +O(\delta^2)$. Also, with the particular ellipsoidal shape and the semi-major axis and semi-minor axis as $ 1 - 5\delta/12$ and $1- 5\delta/6$, respectively, it can be shown that $\Phi$ is linear in $X$; namely, $\Phi/U_{\infty} = (3/2 +3\delta/4)x$. Thus, the slope increases as $\delta = \varepsilon^2$ increases. This is consistent with the exact solutions shown in Figure \ref{potential}.

\begin{figure}[thbp]
\centering
 \includegraphics[height = 7.0cm]{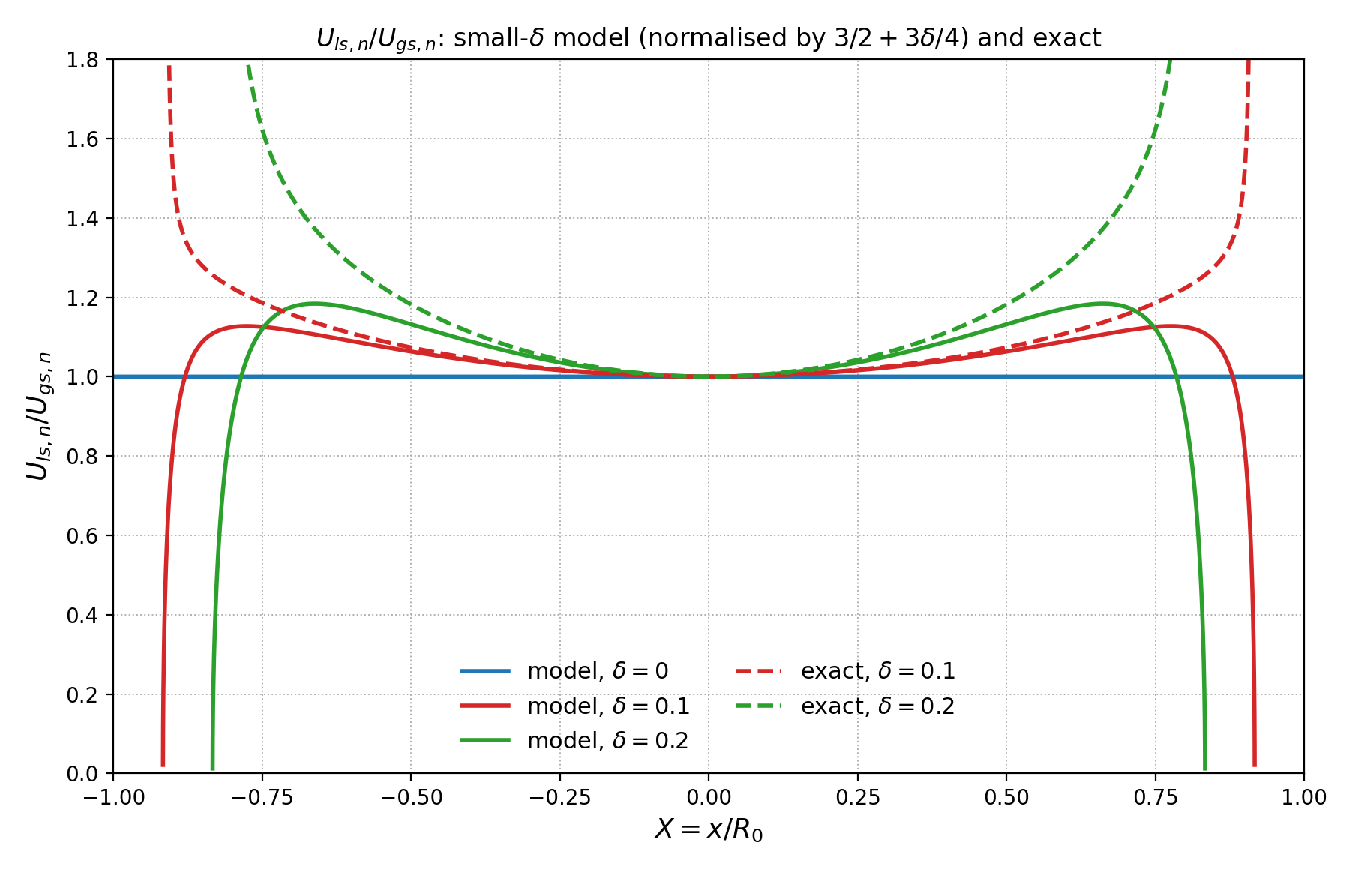}
   \caption{For three values of $\delta = 0, 0.1, 0.2$ , the ratio of liquid surface speed to gas surface speed is plotted along the surface where each speed has first been normalized by its maximum value at $x=0$. The solid lines use perturbation theory for the gas velocity while the dashed lines use the exact theory.}
  \label{velocityratio}
  \end{figure}
In Figure \ref{velocityratio}, we examine the ratio of liquid surface speed to gas surface speed, normalizing so that the ratio has unit value at $x=0$. (Of course, the gas speed  will be much larger.) The blue line for the sphere shows the same profile with a constant ratio  for both phases. For  ellipsoidal droplets, the ratio is not uniform with the liquid speed becoming larger in a relative sense as distance from $x=0$  increases. The solid lines for the perturbation theory show its unreliability near the stagnation points where the velocity magnitudes are small and the higher order terms become relatively more important. 

In order to estimate friction drag, we must consider the viscous boundary lying across both sides of the droplet surface. The droplet surface fluid actually moves at speed $U^*$ where $U_{l,s} < U^* << U_{g,s}$.  For the common fluids previously mentioned, the coefficient for dynamic viscosity can be larger by a forty-to-fifty factor for the liquid than for the gas. However, the density ratio is even larger; thus, even though the liquid velocity is lower, the ratio of the Reynolds number, using the same spherical radius $R_0$ as the length and the maximum surface speeds to define the liquid-phase and gas-phase Reynolds numbers, we have $Re_l/Re_g \approx \sqrt{\rho_l/\rho_g}\mu_g/\mu_l$. Thus, although  $Re_l < Re_g$, they can be in the same magnitude range. Thereby, the boundary layer thickness on both sides can be comparable. We will consider a Reynolds number large enough to consider the layer is thin compared to the droplet  axes. In equating the shear stress on both sides of the droplet surface, the larger liquid viscosity causes the velocity $U^*$ exactly at the interface inside the boundary layer to have a value very close to the liquid velocity at the boundary-layer edge and be substantially different from the gas velocity at the layer edge for these common fluids.

Very limited information exists on drag for the oblate spheroid. \cite{LeClair} numerically determines drag for flow over a sphere with a wide range of $Re$; friction drag and pressure drag are separated and reported in tabular form, but without any attempt at a correlation. \cite{Ouchene}
presents numerical results and correlations for flow over oblate spheroids with $0.1 \leq Re \leq 100$ and  $\lambda$, the ratio of semi-minor axis length to semi-major axis length in the $0.2$ to $1.0$ range. 
For the drag coefficient in the axisymmetric flow configuration (zero angle of attack) the correlation of numerical results gives the drag coefficient as
\begin{eqnarray}
C_D &=& \frac{24}{Re}\Big( K + 0.15 \lambda^{95.91}Re^{0.687} 
+ 0.2927 (1-\lambda)^{0.4374} Re^{0.7512}                     \Big)  \nonumber \\
K &\equiv& \frac{8}{3}\lambda^{-1/3}\Big[ \frac{2\lambda}{1-\lambda^2} + \frac{2(1-2\lambda^2)}{(1 - \lambda^2)^{3/2}} tan^{-1}\big(\frac{\sqrt{1-\lambda^2}}{\lambda}\big)\Big]^{-1}
\end{eqnarray}
The author does not separate pressure drag and friction drag. It is known that at very low Reynolds number with Stokes flow the friction drag coefficient $C_{D,f}$ and pressure drag coefficient $C_{D,p}$ have a two-to-one ratio. Namely,
\begin{eqnarray} 
C_D = C_{D,f} + C_{D,p} = \frac{16}{Re} + \frac{8}{Re}
\end{eqnarray}
 At $Re =20$ both components of the coefficient decrease but pressure drag becomes of comparable importance to friction drag. (Note drag force increases  with increasing velocity for a fixed sphere size.) For our $Re$ range of interest,  the two components are roughly equal for the sphere,  Thus, for our oblate spheroid and our range of $Re$, the friction drag coefficient might be taken  as
 \begin{eqnarray}
 C_{D,f} &=& \frac{12}{Re}\Big( K + 0.15 \lambda^{95.91}Re^{0.687} 
+ 0.2927 (1-\lambda)^{0.4374} Re^{0.7512}                     \Big) 
 \end{eqnarray}
However, the need for more research must be acknowledged on this point.

Based on our analysis, the axes ratio $\lambda = 0.875 + O(We^2)$ for $We = 0.4, \delta =0.1$, while it takes value of approximately $0.75$ for $We =0.8, \delta =0.2$. Consequently, the effect on friction drag due to deformation is significant even for modest values of $We$. Since common gases have Prandtl number $Pr =O(1)$ and Schmidt number $Sc = O(1)$, the impact of modest deformation on heat and mass transport rates and vaporization rate is expected to be large.

\section{Conclusions}  \label{conclude}

Several analyses have been developed and coordinated to obtain a model for the common stable configuration of axisymmetric flow around a translating droplet shaped as an oblate spheroid at low Weber number $We$ and moderate Reynolds number $Re$. The effect of surface-pressure variation and surface tension on shape has been described through a differential analysis with $\delta = We/4$ as the logical perturbation parameter. The internal liquid circulation has been described using an extension of Hill's spherical vortex where the vorticty in the liquid vortex is directly proportional to the cylindrical radial coordinate value and independent of axial position. The liquid velocity is substantially lower than the gas velocity for common configurations because liquid density is orders of magnitude larger.

An image-ring doublet was created and its potential function was added to the potential of a weak doublet at the origin and the free stream to analyze the flow over the oblate spheroid. The doublet-ring radius controlled the ratio of the two axes of the body of revolution. The exact analysis was augmented by a perturbation analysis designed to reduce computational cost. The perturbation parameter was the normalized doublet-ring radius. In general, the two potential-flow analyses agreed well, except near the stagnation points where the $O(\varepsilon^2)$ terms become small and rivaled by higher-order terms that are omitted in the perturbation analysis. The perturbation theory still has value because the errors occur at small radial values affecting a small fraction of the surface area and an even smaller fraction of the liquid volume.

The need for addition of a weak point doublet at the origin to avoid a singularity was identified first through the perturbation analysis but was also followed in the exact analysis. Through coupling with other analyses, $\varepsilon$, the doublet radius normalized by the radius of a spherical droplet of the same volume was set equal to $\sqrt{We}/2$.

There are some ad hoc features of the analysis with modest error. The internal liquid circulation analysis uses an exact inviscid solution for an ellipsoid while surface-tension analysis predicts a spheroidal shape that is near-ellipsoidal for low Weber number $We$ values. The perturbation theory for the gas potential flow predicts an ellipsoidal shape for low $We$, but the exact analysis for the potential flow shows higher-order deviation from an ellipsoid for the spheroidal shape.

Some comments on droplet drag were presented with suggestion for future direction of related work. The model has the ability to be developed further  to include heat and mass transport through both the gas boundary layer and the core of the liquid droplet and to include vaporization with Stefan advection. The modification of transport rates is expected to be significant even for small $We$.

\section*{Appendix: Double-Vortex-Ring Equivalence}

We are able to show an equivalence between the thin circular ring doublet discussed in 
Section \ref{potent} and two thin circular concentric ring vortices of opposite rotation that are brought asymptotically to the same radius $\sigma$. Following  \cite{Fraenkel1970} with appropriate adjustment to our nomenclature. The  ring vortex with vorticity $\omega$ distributed over a circular line centered on the origin in the $x=0$ plane yields a stream function $\Psi_{vr}$ in the following form.
\begin{eqnarray}
    \frac{\Psi}{R_0^2} = \frac{1}{4\pi}\int_0^{2\pi} \frac{r\sigma \omega \cos \alpha\; d\alpha}{\big[r^2 + x^2 + \sigma^2 - 2\sigma r \cos \alpha \big]^{1/2}}
\end{eqnarray}
Next, a second concentric circular ring vortex with radius and vorticity set at $\sigma +\Delta\sigma$ and $-\omega$ (i.e., opposite rotational direction)  is added. Its contribution is evaluated by taking a Taylor series about the $\sigma$ position. The two vortices are summed to yield a stream function $\Psi_{dvr}$ for the double vortex ring. It yields
\begin{eqnarray}
   \frac{1}{R_0^2} \Psi_{dvr} = -\frac{1}{R_0^2}\frac{\partial \Psi}{\partial \sigma}\Delta \sigma &=& -
   \frac{r\omega\Delta\sigma}{4\pi}\int_0^{2\pi}\Big[\frac{\sigma(\sigma - r \cos \alpha)\cos \alpha}{\big[r^2 + x^2 + \sigma^2 - 2\sigma r \cos \alpha \big]^{3/2}} \nonumber \\ && -    \frac{\cos \alpha}{\big[r^2 + x^2 + \sigma^2 - 2\sigma r \cos \alpha \big]^{1/2}}\Big] d\alpha   +O\big(\frac{\omega(\Delta \sigma)^2}{R_0}\big) 
\end{eqnarray}
Next, $\omega \Delta\sigma \rightarrow U_{\infty}/ \sigma$ as $\Delta \sigma \rightarrow 0$ and $\varepsilon = \sigma /R_0$ is substituted. Using the definitions of the definite integrals $I_1$ and $I_4$ in Section \ref{potent} and recognizing a cyclic integrand, we have
\begin{eqnarray}
\frac{\Psi_{dvr}}{U_{\infty}R_0^2} &=& -\frac{r^2}{R_0^2}I_1 +\varepsilon \frac{r}{R_0} I_4 - \frac{R_0^2}{4\pi\sigma} \int_0^{2\pi}  \frac{\partial }{\partial \alpha} \big[ \frac{ \sin \alpha}{\big[r^2 + x^2 + \sigma^2 - 2\sigma r \cos \alpha \big]^{1/2}}\big]
d\alpha  \nonumber \\  & =&  -\frac{r^2}{R_0^2}I_1 +\varepsilon \frac{r}{R_0} I_4 = \frac{\Psi_{ringdoublet}}{U_{\infty}R_0^2}
\end{eqnarray}
Realize that the weak point doublet at the origin plus the free-stream term must be added to match the full solution in Section \ref{potent}.
Furthermore, the double  vortex ring produces a potential flow with the value of circulation set to zero because the two rings have vorticity of equal magnitude but opposite direction. 

\begin{figure}[thbp]
\centering
 \includegraphics[height = 6.0cm]{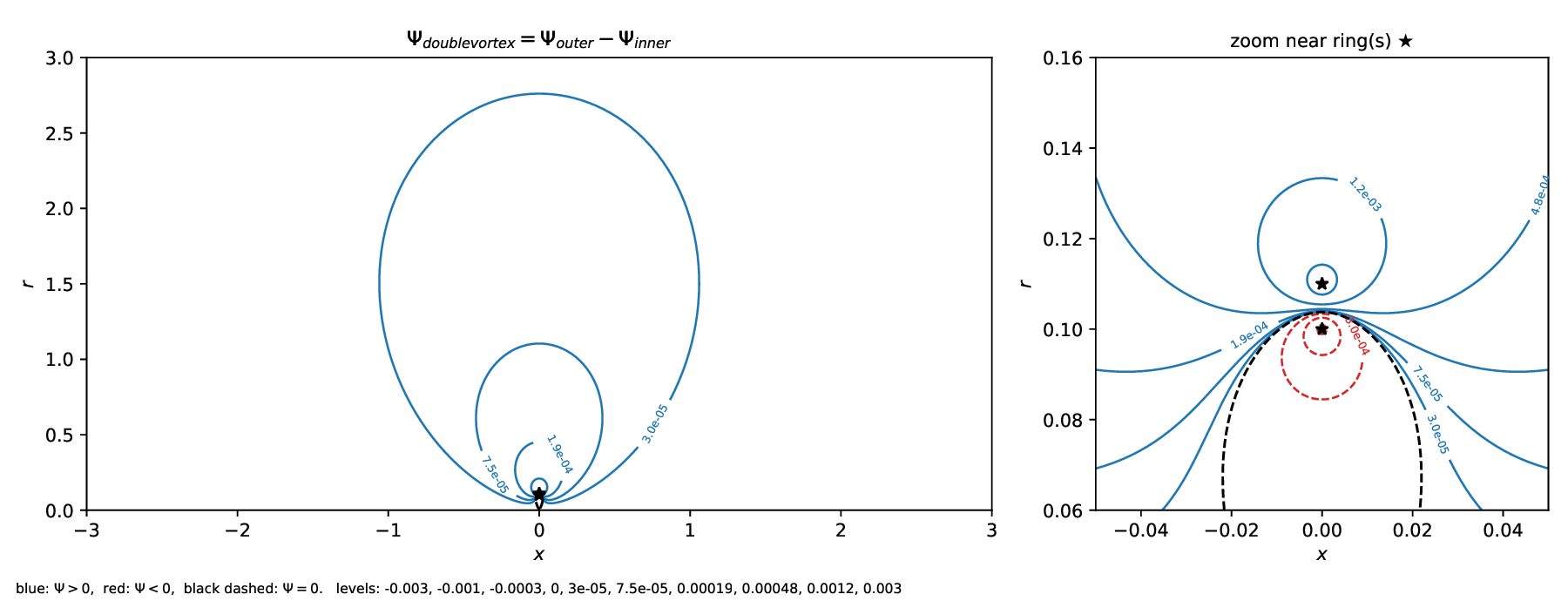}
   \caption{Streamlines for two concentric thin vortex rings of equal vorticity magnitudes but opposite rotation.}
  \label{doublevortex}
  \end{figure}

In Figure \ref{doublevortex}, streamlines are shown for two thin concentric vortex rings of equal vorticity magnitudes but opposite directions of rotation. The two ring  radii are within ten percent of each other. It shows the tendency to behave like a ring doublet, especially with regard to distant points from the rings. It gives insight to how the limiting behavior as the two radii equate is equivalent to the ring doublet.

\section*{Acknowledgement}
 Claude AI was used for numerical calculation and graphing of results shown in the figures.

\bibliographystyle{jfm}
\bibliography{Revised3DCompress.bib}

\end{document}